# One-dimensional van der Waals heterostructures as efficient metal-free oxygen electrocatalysts

Chang Liu[+], Fei Liu[+], Hao Li[+], Junsheng Chen, Jingyuan Fei, Zixun Yu, Ziwen Yuan, Chaojun Wang, Huiling Zheng, Zongwen Liu, Meiying Xu, Graeme Henkelman, Li Wei*, Yuan Chen*

Dr. F. Liu, J. Chen, C. Liu, J. Fei, Z. Yu, Z. Yuan, C. Wang, Prof Dr. Z. Liu, Prof Dr. Y. Chen and Dr. L. Wei

School of Chemical and Biomolecular Engineering

The University of Sydney, Darlington, NSW, Australia, 2006

Email: l.wei@sydney.edu.au (L.Wei) and yuan.chen@sydney.edy.au (Y.Chen)

Dr. F. Liu and Prof Dr. M. Xu

State Key Laboratory of Applied Microbiology Southern China, Guangdong Provincial Key Laboratory of Microbial Culture Collection and Application

Guangdong Institute of Microbiology, Guangzhou, Guangdong, China, 510070

Dr. H. Li, H. Zheng, and Prof G. Henkelman

Department of Chemistry and the Oden Institute for Computational and Engineering Sciences

The University of Texas at Austin, 105 E. 24th Street, Stop A5300, Austin, Texas, United States, 78712

[+] These authors contributed equally.

**Abstract**

Two-dimensional covalent organic frameworks (2D-COFs) are an emerging family of catalytical materials with well-defined molecular structures. The stacking of 2D nanosheets and large intrinsic bandgaps significantly impair their performance. Here, we report coaxial one-dimensional van der Waals heterostructures (1D vdWHs) comprised of a carbon nanotube (CNT) core and a thickness tunable thienothiophene-pyrene COF shell using a solution-based *in-situ* wrapping method. Density functional theory calculations and *in-operando* and *ex-situ* spectroscopic analysis show that the carbon-sulfur region in the thienothiophene groups is the active catalytic site. The unique coaxial structure enables controllable *n*-doping from the CNT core to the COF shell depending on COF shell thickness, which lowers the bandgap and work function of COF. Consequently, the charge transfer barrier between the active catalytic site and adsorbed oxygen intermediates becomes lower, resulting in a dramatic enhancement in their catalytic activity for oxygen redox reactions. It enables a high-performance rechargeable zinc-air battery with a specific capacity of 696 mAh $g_{Zn}^{-1}$ under a high current density of 40 mA $cm^{-2}$ and excellent cycling stability. 1D vdWHs open the door to create multi-dimensional vdWHs for exploring fundamental physics and chemistry, as well as practical applications in electrochemistry, electronics, photonics, and beyond.
.

Two-dimensional covalent organic frameworks (2D-COFs) are planar nanosheets in a periodic lattice, assembled from covalently bonded small molecules.[1, 2] They possess several outstanding properties for enabling high activity catalysts, for example, a large specific surface area, tunable porosity, and excellent thermal and chemical stability. More importantly, versatile molecular building blocks of 2D-COFs may offer well-defined and structurally tunable catalytic active sites with atomic precision.[3] Recently, 2D-COFs have been explored as a new family of emerging electrocatalysts for the oxygen reduction reaction (ORR) and oxygen evolution reaction (OER), which are required in many energy storage and conversion devices, such as zinc-air batteries (ZABs), fuel cells and water electrolyzers.[4] However, there are two challenges for these materials to serve as efficient electrocatalysts. First, abundant out-of-plane π-electrons on 2D-COFs cause spontaneous stacking of 2D nanosheets, resulting in thick COF laminates, blocking the access to surface-active catalytic sites and sabotaging mass transfer of electrolytes and reaction intermediates.[2] Second, their conjugated π-electron skeletons may open few-eV intrinsic bandgaps, impairing efficient electron transfer.[5]

Creating nanohybrids of 2D-COFs and carbon nanotubes (CNTs) may overcome the above challenges because CNTs can not only serve as spacers to prevent the stacking of 2D-COF, but also provide fast electron transfer paths. Along these lines, CNTs have already been used to synthesize composites with 2D graphene or g-$C_3N_4$ nanosheets.[6, 7, 8] Further, intermolecular charge-transfer may take place between CNTs and their surface adsorbates, which have been explored to modulate the catalytic activity of the adsorbates.[8-10] Composites can be formed by metallic, ionic, or covalent bonds. However, precise control of these strong chemical bonds is difficult, and the formation of these strong chemical bonds can also significantly interrupt the intrinsic properties of constituting components. Alternatively, 2D crystals have been assembled into van der Waals heterostructures (vdWHs) by stacking multiple 2D layers on top of each other, which represents a new way to form crystals.[11] In a recent study, 1D vdWHs were synthesized by growing boron nitride or molybdenum disulfide crystals on CNTs.[12] We envision that coaxial 1D vdWHs can be synthesized via π-electron interactions between 2D-COF and CNTs via bottom-up synthesis, in which 2D-COFs may serve as the shell, while CNTs are the core. The unique interactions between 2D-COFs and CNTs in such 1D vdWHs may enable high-performance electrocatalysts.

Herein, we report the design and synthesis of novel 1D vdWHs comprised of 2D-COFs and CNTs. We used an *in-situ* wrapping method to grow and assemble a catalytically active thienothiophene-containing 2D-COF shell around a catalytic inert multi-walled CNT (MWCNT) core. This method allows the precise tuning of the thickness of the COF shell in the nanometer scale, leading to the discovery of the shell thickness-dependent high electrocatalytic activity for both ORR and OER. Density functional theory (DFT) calculations, diffusive reflectance ultraviolet-visible spectroscopy (UV-vis DRS), and ultraviolet photoelectron spectroscopy (UPS) were applied to understand electronic interactions in the 1D vdWHs, revealing significant *n*-doping from CNTs to COF shells. Further theoretical calculations and *in-operando* Fourier transformed infrared spectroscopy (FTIR) studies were used to identify the specific catalytic active site. The practical application of the 1D vdWH as a highly efficient bifunctional oxygen electrocatalyst was also demonstrated in high-performance rechargeable ZABs.

As illustrated in **Figure 1a**, a thienothiophene containing molecule, *thieno[3,2-b]thiophene-2,5-dicarboxaldehyde* (**TtDCA**), was used to construct a 2D-COF with *4,4',4'',4'''-(pyrene-1,3,6,8-tetrayl)tetra-aniline* (**TTAP**). We chose TtDCA because the two symmetric S-pentacyclic domains in its thienothiophene group can readily mimic the S-C structures in S doped carbon materials, which can act as bifunctional catalytic sites for ORR and OER.[13] The Schiff-base condensation between TtDCA and TTAP in the presence of CNTs in *N, N*-dimethylacetamide (DMAc) produced 1D coaxial vdWHs with COF shell of a tunable thickness (see Experimental Section for details). The thickness of the COF shell can be controlled by tuning the mass ratio between CNTs and COF precursors ($m_{CNT}$: $m_{precursors}$ = 1, 2, or 4). The resulting 1D vdWHs are denoted as CC-X, where X refers to the thickness of the COF shell in nanometers measured by transmission electron microscope (TEM).

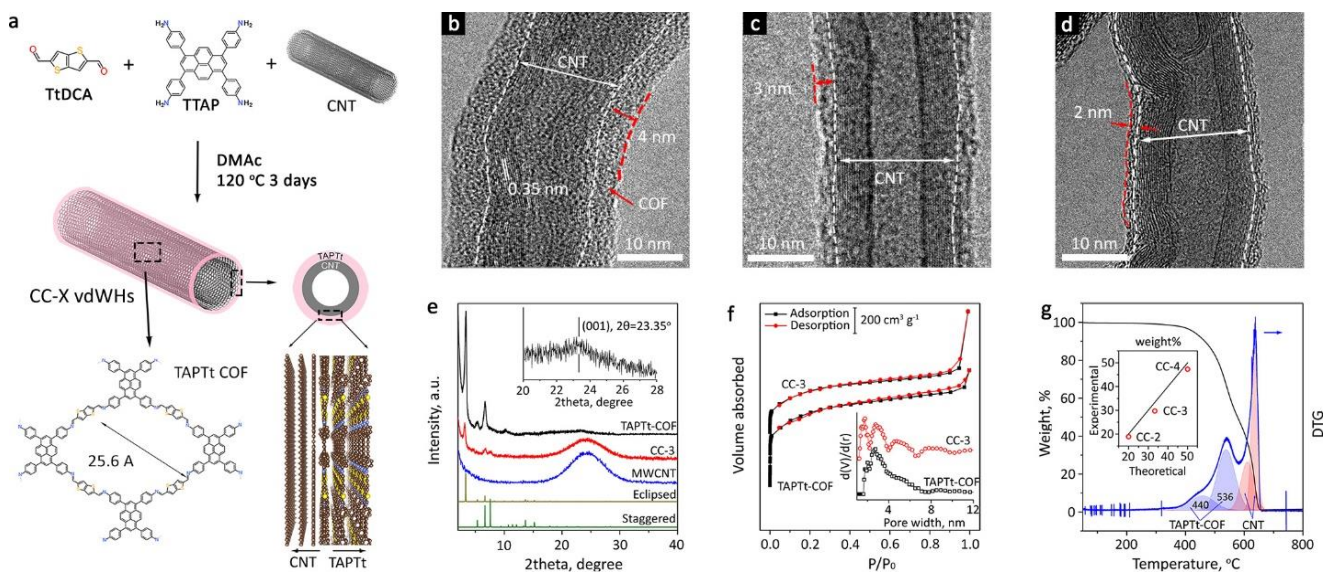

**Figure 1.** a) Schematic illustration of the synthesis of 1D vdWHs (CC-X). b-d) TEM images of (b) CC-4, (c) CC-3, and (d) CC-2. e) XRD patterns (inset: magnified (001) region of TAPTt-COF). f) $N_2$ physisorption isotherms of TAPTt-COF and CC-3 (inset: pore size distributions). g) TGA and DTG profiles of CC-3 (inset: weight fraction of COF in CC-X).

The successful formation of TAPTt-COF was confirmed by FTIR and $C^{13}$ solid-state nuclear magnetic resonance (ssNMR), as displayed in Figure S1 and S2 in the Supporting Information (SI). An atomic force microscope (AFM) image and the corresponding height profiles of pristine TAPTt-COF (Figure S3 in SI) exhibits a large nanosheet (~2 × 4 $\mu m^2$) with a thickness of about 1.5 nm (4 atomic layers), confirming its 2D structure. Small diameter multi-walled CNTs (MWCNTs) used in this study have smooth and clean surfaces (see TEM images in Figure S4 in SI). In comparison, TEM images in Figure 1b-d show that seamless COF shells with varied thicknesses are formed on CNTs. The thickness of the COF shells can be precisely controlled at a nanometer-scale precision. Figure 1b-d display the thickness of COF shells is 4.2, 3.0, and 1.9 nm in the 1D vdWHs synthesized using CNTs to COF precursors at the mass ratio ranging from 4, 2, to 1 (**Table 1**).

Strong π-π interactions take place between CNTs and pyrene groups in TTAP,[14] and we expect that the interactions play a critical role in guiding the formation of the 1D core-shell structured vdWHs.

Without the presence of CNTs, pristine TAPTt-COF self-assemblies into microspheres with a diameter of about 2~3 $\mu$m, as shown in the SEM images in Figure S5 in SI. These COF microspheres contain well-aligned nano-channels around 2.5 nm wide, as confirmed by TEM observation shown in Figure S6 in SI, consistent with a molecular model of this type of COF. In contrast, only tubular structures are observed in the SEM images of CC-X vdWHs (Figure S7 in SI). Although the morphology of the COF shells changes significantly by the CNT cores, they retain their chemical characteristics. Figure S8 in the SI shows that their FTIR spectra are similar to that of pristine TAPTt-COF. Their X-ray photoelectron spectroscopy (XPS) survey scans (Figure S9 in SI) and the corresponding high-resolution XPS spectra of C1$s$, N1$s$, and S2$p$ are virtually identical to that of pristine TAPTt-COF (Figure S10-S12 in SI). Figure S13 in the SI shows that O exists in chemisorbed water on CNTs and CC-X vdWHs with a low O content (<3 at.%). It should be noted that the XPS results indicate a low concentration of oxygenated functional groups, which were considered as active catalytic sites in a previous study.[15]

**Table 1.** Physiochemical properties of TAPTt-COF, CNTs, and CC-X 1D vdWHs.

| Samples | $m_{CNT}$: $m_{precursors}$ | Thickness of COF shell [a], nm | SSA, m$^2$ g$^{-1}$ | COF mass loading [b], % |
|---|---|---|---|---|
| CC-4 | 1 | 4.2 ± 0.5 | 629 | 47.3 |
| CC-3 | 2 | 3.0 ± 0.4 | 436 | 29.7 |
| CC-2 | 4 | 1.9 ± 0.2 | 296 | 18.9 |
| TAPTt | - | - | 1059 | - |
| MWCNT | - | - | 112 | - |

[a] The thickness was measured by TEM. [b] The mass loading was determined by DTG.

Powder X-ray diffraction (XRD) patterns of pristine TAPTt-COF, CC-3, and CNTs are presented in Figure 1e. TAPTt-COF shows an XRD pattern consistent with that of its simulated eclipsed stacking model (Figure S14 in SI). It has an interlayer $d$-spacing of 0.36 nm from the most intense (001) reflection at a *2θ* of 23.35º (inset of Figure 1e, Cu Kα, λ=1.5406 Å). In comparison, the CC-3 VdWH displays not only the features of TAPTt-COF but also a broad peak from the (001) diffraction of CNTs at ~25º. A new peak emerges at 7.59º, which can be assigned to the (111) reflection of TAPTt COF in a staggered stacking model (Figure S15 in SI). $N_2$ physisorption isotherms and corresponding pore size distribution profiles are displayed in Figure 1f and Figure S16 in the SI. TAPTt-COF exhibits a large specific surface area of 1059 $m^2 g^{-1}$ with abundant mesopores at ~2.5 nm. The specific surface area of CC-X vdWHs is listed in Table 1, showing a decreasing trend from 629 to 296 $m^2 g^{-1}$ with reducing COF shell thickness. Their pore size distribution profiles (Figure S16 in the SI) show a new pore at ~1.6 nm, which can be assigned to such pores in TAPTt COF packed in the staggered stacking model (Figure S15 in the SI).

The mass loading of TAPTt-COF in CC-X vdWHs was determined by thermogravimetric analysis (TGA) and differentiated thermogravimetric analysis (DTG). Figure 1g shows the representative TGA and DTG profiles of CC-3. The deconvolution of the DTG profile reveals two distinctive groups of peaks at 440 and 536 ºC and >600 ºC, which can be attributed to the thermal decomposition of TAPTt-COF and CNTs (Figure S17 in SI), respectively. The mass loading of COF in CC-3 is 29.7 wt% (Table 1), which is similar to the mass ratio between CNT and COF precursors (2: 1). The mass loadings of COF in CC-4 and CC-2 exhibit the same agreement (inset of Figure 1g and Figure S17 in the SI). Overall, different physiochemical characterization results confirm that our *in-situ* wrapping method can precisely control the thickness of COF shells in the resulting 1D core-shell structured vdWHs.

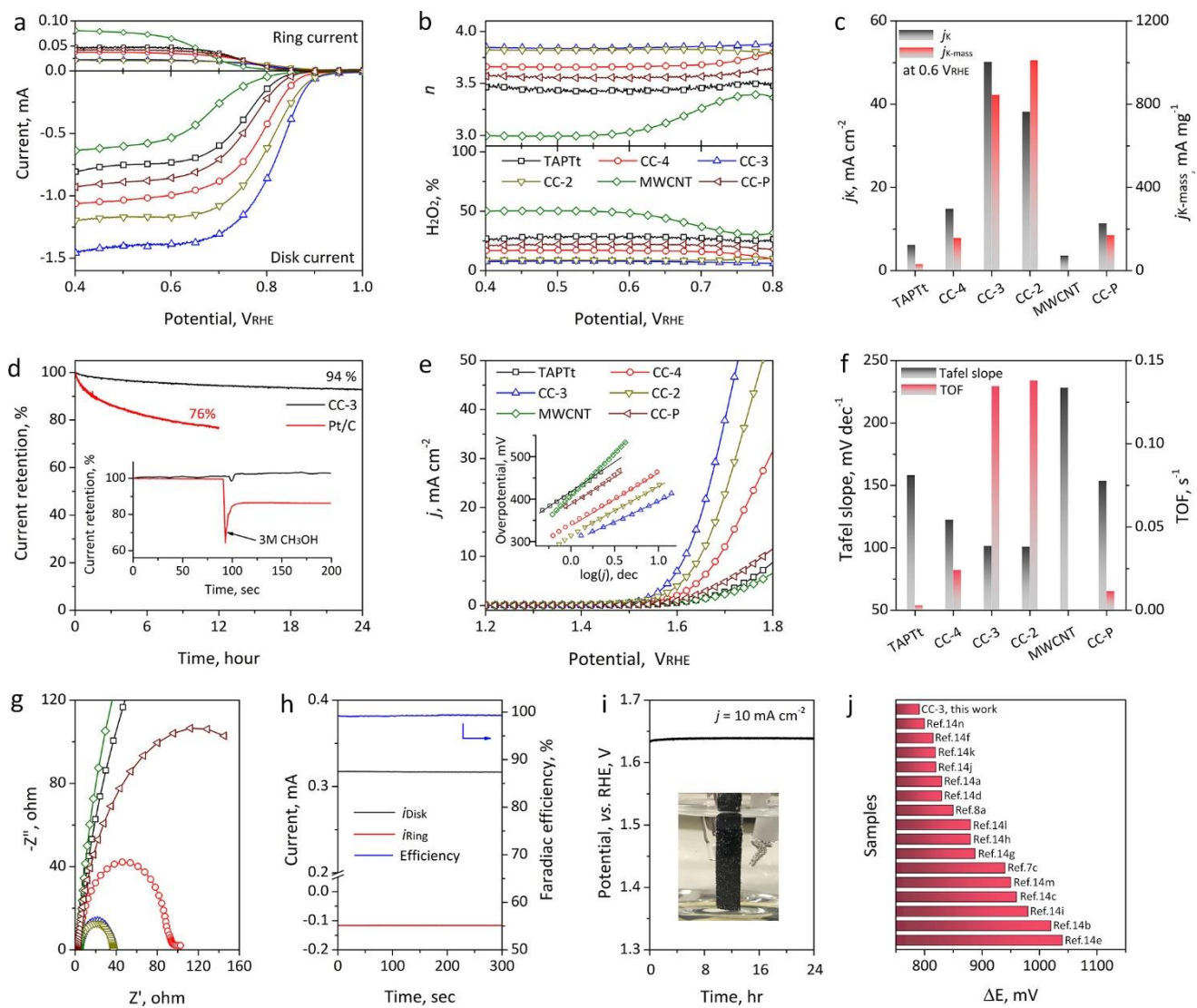

**Figure 2.** Electrocatalytic performance. a) ORR RRDE-LSV curves and b) calculated $n$ and $H_2O_2$%. c) Comparison of $n$ and $j_K$ at 0.6 $V_{RHE}$. d) ORR stability and methanol tolerance test (inset) of CC-3. e) OER-LSV curves and Tafel plots (inset). f) Comparison of Tafel slopes and TOFs. g) EIS Nyquist plots. h) Faradaic efficiency and i) OER stability of CC-3. j) Comparison of recently reported bifunctional oxygen electrocatalysts with CC-3.

The electrocatalytic performances of CC-X vdWHs, together with several commercial catalysts as references, were evaluated in $O_2$ saturated 0.1 M KOH electrolyte. **Figure 2a** shows their ORR linear sweep voltammetry (LSV) curves collected on a rotary ring-disk electrode (RRDE). Their half-wave potentials ($E_{1/2}$) are listed in **Table 2** for comparison. There is a clear trend that the electrocatalytic activity of CC-X vdWHs depends on the thickness of their COF shells. CC-3 exhibits the highest activity for ORR activity among the three vdWHs with the highest $E_{1/2}$ of 0.828 V (*vs.* the reversible hydrogen electrode, RHE), which is also much higher than that of pristine TAPTt-COF at 0.741 V, CNTs at 0.698 V, and a physical mixture of TAPTt-COF and CNT at 0.761 V (at a mass ratio of 1:2, denoted as CC-P, see Experimental Section for details). The electron transfer number (*n*) and the selectivity to $H_2O_2$ selectivity are calculated and compared in Figure 2b. CC-X vdWHs follow a stable $4e^-$ ORR pathway in the tested potential window. The *n* of CC-3 is ~3.86, close to that of 20 wt.% Pt/C at 3.95 (Figure S18 in SI). Further, Koutecky–Levich (K-L) plots of different samples are compared to envisage their kinetic properties (Figure S19 in SI). The electron transfer numbers were also determined from the K-L plots at 0.7 $V_{RHE}$, and the results are listed in Table 2.

**Table 2.** Electrocatalytic performances for ORR and OER of different catalysts.

|  | ORR performance | | | | | OER performance | | | |
|---|---|---|---|---|---|---|---|---|---|
|  | $E_{1/2}$, V | $n^{\,a}$ | $n^{\,b}$ | $j_K^{\,b}$, mA $cm^{-2}$ | $j_{K\text{-mass}}^{\,b}$, A $mg^{-1}$ | $\eta_{10}$, mV | Tafel slope, mV $dec^{-1}$ | TOF$^{\,c}$, $s^{-1}$ | $R_{CT}^{\,d}$, ohm |
| CC-4 | 0.771 | 3.62 | 3.64 | 14.86 | 0.16 | 457 | 122 | 0.0242 | 98.6 |
| CC-3 | 0.828 | 3.86 | 3.89 | 50.17 | 0.84 | 389 | 101 | 0.134 | 36.2 |
| CC-2 | 0.799 | 3.90 | 3.91 | 38.19 | 1.01 | 409 | 100 | 0.138 | 35.4 |
| TAPTt | 0.741 | 3.45 | 3.42 | 6.21 | 0.03 | 605 | 158 | 0.00288 | 486.9 |
| MWCNT | 0.698 | 3.21 | 3.14 | 3.60 | - | 751 | 228 | - | 551.2 |
| CC-P | 0.761 | 3.54 | 3.57 | 8.37 | 0.13 | 578 | 153 | 0.0114 | 223.7 |
| Pt/C | 0.865 | 3.95 | 3.96 | 46.57 | 1.16 | - | - | - | - |

*a.* Determined from RRDE tests at 0.7 $V_{RHE}$; *b.* Calculated from K-L plots at 0.7 $V_{RHE}$; *c.* Calculated at $\eta = 400$ mV; *d* Obtained at $\eta = 350$ mV

The kinetic current densities ($j_K$) at 0.6 $V_{RHE}$ of different samples are compared in Figure 2c. The $j_K$ of CC-3 (50.17 mA $cm^{-2}$) and CC-2 (38.19 mA $cm^{-2}$) is much higher than that of CC-1 (14.86 mA $cm^{-2}$) and CC-P sample (9.37 mA $cm^{-2}$). Besides, the $j_K$ of CC-3 is also higher than that of 20 wt.% Pt/C (46.57 mA $cm^{-2}$). Assuming TAPTt COF or Pt as the active catalytic component, the mass-normalized kinetic current density ($j_{K\text{-mass}}$) was calculated and listed in Table 2. Figure 2c shows that the $j_{K\text{-mass}}$ decreases with an increase in the COF shell thickness. CC-2 has the largest $j_{K\text{-mass}}$ of 1.01 A

mg$^{-1}$, which is ~16% higher than that of CC-3 (0.84 A mg$^{-1}$) and over 30 times larger than that of TAPTt-COF (0.03 A mg$^{-1}$). However, because the mass loading of COF in CC-2 is lower than that in CC-3, CC-3 exhibits better overall catalytic performance for ORR. The stability test results shown in Figure 2d indicate that CC-3 has better stability than Pt/C. It retains 94% of its initial current density after the 24-h chronoamperometric test at 0.4 V$_{RHE}$. In comparison, Pt/C lost 24% of its initial current density in 12 h. CC-S also exhibits good methanol tolerance (inset of Figure 2d).

The OER performance was also assessed in 0.1 M KOH electrolyte. Figure 2e displays LSV curves of different samples. TAPTt-COF, CNTs, and CC-P show negligible OER catalytic activity. In contrast, CC-X vdWHs have much higher activities (Table 2). Among the three samples, CC-3 demonstrates the best catalytic performance with the lowest overpotential ($\eta_{10}$) of 389 mV to achieve the current density ($j$) of 10 mA cm$^{-2}$. The inset of Figure 2e shows Tafel plots of different samples, and their Tafel slopes are compared in Figure 2f. CC-3 and CC-2 have comparable Tafel slopes of 101 and 100 mV dec$^{-1}$, respectively, indicating similar OER kinetics.

The current densities of different samples were further normalized according to their electrochemically active surface area (ECSA, measured by the cyclic voltammetry (CV) scanning method, Figure S20 in SI). The $j_{ECSA}$ (Figure S21 in SI) suggests a similar performance trend as that in Figure 2d, except that the current density difference between CC-3 and CC-2 is reduced. Assuming that the active catalytic sites are on TAPTt-COF, we calculated turnover frequencies (TOF) of our different samples, as shown in Figure S22 in the SI (see the Experimental Section for details). The TOFs at a $\eta$ = 400 mV are compared in Figure 2f. The TOF of CC-2 at 0.138 s$^{-1}$ is slightly higher than that of CC-3 at 0.134 s$^{-1}$, suggesting its higher intrinsic activity, which is consistent with the results of their catalytic activity for ORR.

Electrochemical impedance spectroscopy (EIS) was used to determine the charge transfer resistance ($R_{ct}$). A Nyquist plot is shown in Figure 2g, and the values of $R_c$ determined at $\eta$ = 350 mV are listed in Table 2. $R_{ct}$ depends on the thickness of the COF shells. Figure 2h shows that CC-3 delivers a near-unity Faradaic efficiency (99.2%), as determined by an RRDE method.[16] It also displays excellent stability with less than 4% overpotential increment in the 24-h chronopotentiometric stability test under 10 mA cm$^{-2}$ (Figure 2i). The inset of Figure 2i shows a photo of oxygen bubbles on a CC-3 electrode.

We further compared ORR $E_{1/2}$ and OER $\eta_{10}$ of CC-3 with those of recently reported metal-free and carbon-based bifunctional oxygen electrocatalysts (listed in Table S1 in the SI). The potential difference ($\Delta E$) between the $E_{1/2}$ and $\eta_{10}$ collected in 0.1 M KOH electrolyte is calculated to compare their bifunctional catalytic activity. As displayed in Figure 2j, CC-3 shows the smallest $\Delta E$ of 791 mV. Alternatively, we also compiled a catalytic activity atlas by plotting ORR $E_{1/2}$ against OER $\eta_{10}$. As shown in Figure S23 in the SI, CC-3 is located at the bottom-left, suggesting one of the best catalytic performances among recently reported metal-free carbon electrocatalysts.[7, 9, 17]

Electrocatalytic performance test results above suggest that the ORR and OER activity of 1D CC-X vdWHs depends strongly on the thickness of COF shells. We combined theoretical and spectroscopic studies to understand this dependence, as well as to explore the origin of their superior catalytic activity. We first analyzed electronic interactions by DFT calculations. The atomic geometry optimized CC-X vdWHs was modeled as thienothiophene moieties sitting on a graphene substrate. **Figure 3a** shows that the graphene substrate (electron-deficient, blue color) injects abundant delocalized electrons to thienothiophene moieties (electron-rich, yellow color), resulting in the *n*-doping effect to the COF shell. The delocalized electrons, which can be treated as excessive carriers, can change the COF shell electronic properties, and consequently, its electrochemical activity.[18]

We hypothesize the excessive carriers in the COF shell should follow the well-established carrier continuity equation, which describes the decaying of excessive carriers in semiconductors as schematically illustrated in Figure S24 in SI.[19] To prove our hypothesis, we compared the near-surface carrier density of CC-X vdWHs and pristine TAPTt-COF by applying the Mott-Schottky analysis.[20] As shown in Figure S25 in SI, their Mott-Schottky plots (reciprocal square root of capacitance *vs.* potential) all exhibit positive slopes, confirming the *n*-doping in the CC-X vdWHs.[21]

The value of this slope is inversely proportional to the near-surface carrier density. The smallest slope of CC-2 indicates its highest carrier density among the CC-X vdWHs (see Table 2). We further plotted the reciprocal of the Mott-Schottky slopes as a function of the COF shell thickness in CC-X vdWHs in Figure S26. The fast decaying trajectory can be well fitted to the solution of carrier continuity equation at steady state.[19] Thus, the COF thickness-dependent catalytic activity can be explained by the weakened electronic interaction between COF and CNT when the COF-shell thickness increases.

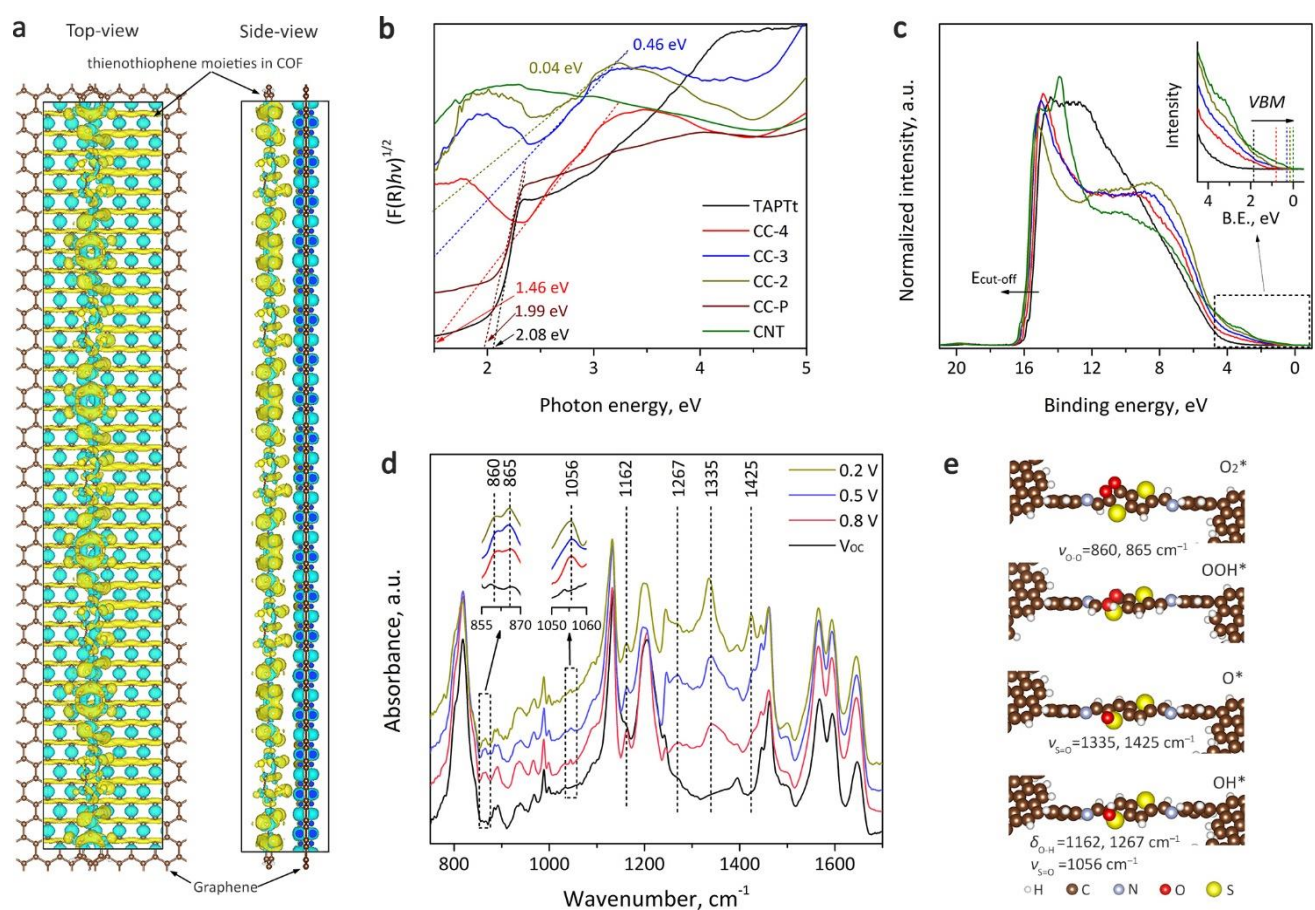

**Figure 3.** Mechanism investigation by theoretical calculations and spectroscopic studies. a) Calculated charge density difference profiles of CC-X vdWHs. b) DRS and c) UPS spectra (inset: spectra near the Fermi level). d) *In-operando* FTIR spectra of CC-3 in $O_2$ saturated 0.1 M KOH electrolyte under different ORR potentials. e) DFT optimized intermediate adsorption geometries and the corresponding experimental IR band assignments.

We further probed the *n*-doping effects in CC-X vdWHs by measuring their indirect bandgap using UV-vis DRS and near-surface work function ($\Phi$) using UPS. Figure 3b shows that the bandgap of CC-X vdWHs decreases with the reduction of the COF shell thickness. TAPTt-COF has a bandgap of 2.08 eV, and CC-4, CC-3, and CC-2 have a bandgap of 1.58, 0.46, and 0.04 eV, respectively, approaching the bandgap of CNTs (0 eV, Table 3). Importantly, the physically mixed CC-P has a bandgap of 1.99 eV, similar to that of TAPTt-COF, indicating that the bandgap modulation by CNTs is only possible when there are strong interactions between COF shells and CNTs. Further, the change of $\Phi$ measured by UPS confirms the same trend (Table 3). USP spectra in Figure 3c show that the valence band maximum (VBM) upshift closer to the Fermi level in CC-X vdWHs with thinner COF shells as a result of higher delocalized electron density.

**Table 3.** Electronic properties of TAPTt COF, CNT, and CC-X vdWHs.

| Sample | Bandgap, eV | $\Phi$, eV | M-S slope, $\times 10^8\,F^{-2}\,V^{-1}$ |
|---|---|---|---|
| TAPTt | 2.08 | 5.47 | 1.29 |
| CC-4 | 1.58 | 5.21 | 1.11 |
| CC-3 | 0.46 | 4.99 | 0.79 |
| CC-2 | 0.04 | 4.86 | 0.68 |
| CNT | ~0 | 4.76 | - |

Based on the above results, we propose that the superior catalytic activity of CC-X vdWHs is related to the *n*-type electronic interaction between CNTs and COF shells. The *n*-doping lowers the bandgap and work function of COF shells, leading to reduced charge transfer barrier between COF shells and adsorbed oxygen intermediates.[22] Further, CC-2, with the thinnest COF shell, experiences the most substantial *n*-doping, thus exhibiting the highest catalytic activity for ORR and OER. However, CC-3 contains a higher mass ratio of COF, which translates into a significantly increased active site number and better catalytic performance. Overall, CC-3 was identified as the optimal catalyst among the three CC-X vdWHs.

We further carried out theoretical calculations and spectroscopic studies to confirm the above-proposed origin of catalytic activity and elucidate the active catalytic center for oxygen redox reactions. Using the elementary steps and parameters proposed for ORR in previous studies[23, 24], we calculated the free energy diagram of ORR intermediates adsorbed on possible active sites of TAPTt COF. All atoms in the proposed model were allowed to relax. DFT results (Figure S27 in the SI) show that the C-S region on the thienothiophene moiety is the only possible active site. The initial $O_2$ adsorption preferentially takes place at a C atom neighboring an S atom in the thienothiophene ring. After accepting a proton-electron pair, the OOH* intermediate diffuses to the S atom and proceeds along the dissociative ORR pathway. Figure S28 in the SI shows that all other atoms are either inert for the initial $O_2$ adsorption or have weak adsorbate binding properties, ruling out their possibility as active sites.

Next, we utilized *in-operando* FTIR to verify the proposed active catalytic center on CC-3 during ORR. Figure 3d shows the *in-operando* FTIR spectra collected under open circuit potential ($V_{OC}$) and different ORR potentials. The IR bands associated with the structure of TAPTt-COF are mostly unchanged, confirming its excellent electrochemical stability. When different ORR potentials were applied, several new features emerge, which are associated with varying intermediates of reaction, as displayed in Figure 3e. The peaks at 860 and 865 cm$^{-1}$ can be assigned to the O–O stretching ($v_{O-O}$) in $O_2$* or OOH* intermediates.[25] The S-O stretching ($v_{S-O}$) peak at 1056 cm$^{-1}$ is close to our DFT-calculated frequency of 1066 cm$^{-1}$. The O-H bending in S-O-H ($\delta_{O-H}$) at 1162 and 1267 cm$^{-1}$ and the S=O stretching mode ($v_{S=O}$) at 1335 and 1425 cm$^{-1}$ originate from intermediates formed along the dissociative ORR pathway. Further, the intensity of these S-O peaks increases under lower applied potentials, which is consistent with the reaction pathway determined from our DFT calculations.[26] Thus, both theoretical calculations and spectroscopic studies confirm that the pre-designated C-S region on the thienothiophene moiety of TAPTt COF is the active catalytic center for oxygen redox reactions.

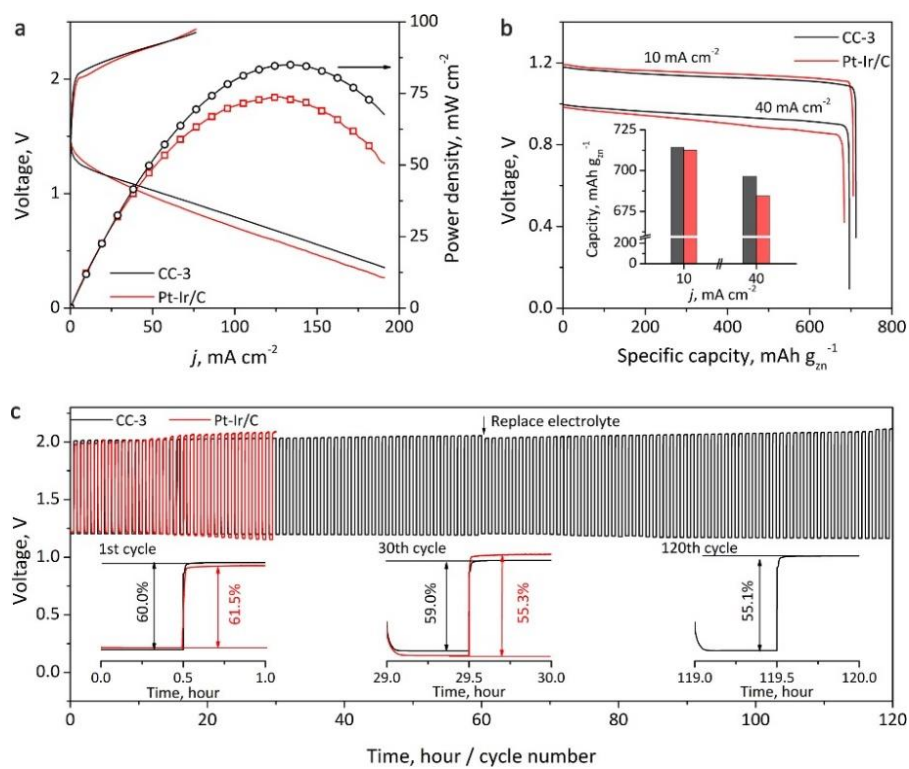

**Figure 4.** Performance of rechargeable zinc-air batteries assembled using CC-3 and Pt-Ir/C oxygen catalysts. (a) Galvanodynamic charge/discharge profiles and power density and (b) galvanostatic discharge curves. The inset shows the battery capacity under different discharge current densities. (c) Cycling profiles. The insets show cycling performances at 1st, 30th, and 120th cycles.

We further demonstrated the application of CC-3 as an efficient bifunctional $O_2$ electrocatalyst to enable high-performance rechargeable ZABs. The details of battery assembly are described in the Experimental Section. Figure S29 in SI shows that the CC-3 ZAB has an open circuit potential of 1.477 V, which is comparable to 1.489 V of the ZAB assembled using commercial Pt/C and $IrO_x$/C catalysts (denoted as Pt-Ir/C). **Figure 4a** shows galvanodynamic charge/discharge profiles of two ZABs. The CC-3 ZAB delivers a maximum power density of 85 mW cm$^{-2}$, which is 21% higher than that of the Pt-Ir/C ZAB (74 mW cm$^{-2}$). Under the discharging current density of 10 mA cm$^{-2}$, the capacity of the CC-3 ZAB is 714 mAh $g_{Zn}^{-1}$, comparable to 712 mAh $g_{Zn}^{-1}$ of the Pt-Ir/C ZAB (Figure 4b). Under the high discharging current density of 40 mA cm$^{-2}$, the CC-3 ZAB outperforms the Pt-Ir/C ZAB (696 *vs.* 684 mAh g$_{Zn}^{-1}$, see the Inset of Figure 4b).

The ZAB cycling performance was evaluated by 120 discharging/charging cycles ($j$ =10 mA cm$^{-2}$, charge/discharge capacity is 5 mAh cm$^{-2}$ per cycle, Figure 4c). Both ZABs show similar performance at the first cycle with a comparable round-trip efficiency of 60.0 and 61.5%, respectively (see the inset on the left in Figure 4c). After 30 cycles, the efficiency of the CC-3 ZAB declines slightly to 59.0%, while that of the Pt-Ir/C ZAB quickly drops to 55.3%. After 120 cycles, the CC-3 ZAB still retains a high efficiency of 55.1%. We also recorded *ex-situ* XPS spectra of CC-3 after the 120-cycle rechargeability test. Figure S30 in SI shows that there are negligible changes in N1$s$ spectra, indicating the high electrochemical stability of imine bonds in TAPTt COF. In contrast, C1$s$, O1$s,$ and S2$p$ spectra exhibit substantial changes. Intensified C-O and C=O features are found in the deconvoluted C1s spectrum. The O1$s$ peak also shifts to the lower binding energy of 531.6 eV, which can be attributed to the formation of C-O and S-O chemical bonds. A new peak emerged in the S2$p$ spectrum at 168.8 eV, which can be assigned to S-O chemical bonds, confirming the adsorption of O intermediates on S.[27] These XPS results are consistent with our DFT calculations and *in-operando* FTIR results, further establishing the C-S region in thienothiophene as the active catalytic center for oxygen redox reactions.

In summary, we report the first metal-free and carbon-based coaxial 1D vdWH, which has a CNT core and TAPTt-COF shells with a tunable thickness from 2 to 4 nm. At the COF shell thickness of 3 nm, the optimal CC-3 vdWH can catalyze the OER via the 4e$^-$ pathway to deliver a TOF of 0.134 s$^{-1}$ at an overpotential of 400 mV. The small potential difference between ORR E$_{1/2}$ and OER $\eta_{10}$ of 791 mV outperforms most recently reported metal-free bifunctional oxygen electrocatalysts. It enables high-performance rechargeable ZABs with a remarkable specific capacity of 696 mAh g$_{Zn}^{-1}$($j$ = 40 mA cm$^{-2}$) and excellent cycling stability. Using DFT calculation, UV-vis DRS, and UPS analysis, we show that the strong *n*-type electronic interaction between CNTs and COF shells is the origin of the significantly improved catalytic activity. Further, we provide, for the first time, a quantitative explanation to the COF shell thickness-dependent catalytic activity by the carrier decay model, providing mechanistic insights into the function of such hetero-structured catalysts. The well-defined chemical structure further pinpoints the C-S region on the thienothiophene moiety of TAPTt COF as the active catalytic center through our theoretical calculations, experimental *in-operando* FTIR, and *ex-situ* XPS studies. Our synthesis strategy demonstrated here opens a new avenue to design and synthesize various multi-dimensional vdWHs for exploring fundamental physics and chemistry, as well as practical applications in electrochemistry, electronics, photonics, and beyond.

**Experimental Section**

*Material synthesis*: MWCNTs (FT9000, CNano, 10-25 nm in diameter) were first purified by a 2-h thermal treatment at 300 °C in airflow, followed by refluxing in 3 M HCl for 6 h to remove amorphous carbon and metal residues. Purified CNTs were washed by deionized water and then dried in a vacuum oven. Afterward, CNTs were thermally annealed in Ar flow (99.999%, BOC) at 1000 °C for 2 h to remove surface functional groups, and then dispersed in DMAc by bath sonication at a concentration of 5 mg mL$^{-1}$. TTAP and TTDCA were dissolved in 5 mL of CNT dispersion at a molar ratio of 1: 2. Assuming the production yield of COF at around 80%, the quantities of TTAP and TTDCA were adjusted to yield CC-X 1D vdWHs with different weight ratios between CNTs and COF precursors ($m_{CNTs}$: $m_{precursors}$ = 1, 2 and 4). The mixtures were frozen and thawed for three cycles before adding 10 μL of 6 M aqueous acetic acid solution and flamed sealed. The solvothermal reaction took place at 120 °C and last for 3 days under stirring. After cooling to room temperature, the solid products were recovered by filtration and repeatedly washed with pure DMAC before dried under vacuum. The reference CC-P sample was prepared by physically mixing TAPTt-COF and CNTs at a mass ratio of 1: 2. The mixture was ground in an agate mortar for 30 min and then dissolved in isopropanol. The solvent was evaporated after bath sonication for 60 min.

*Characterization*: The FTIR spectra were collected on an FTIR spectrometer (Nicolet 6700, Thermo Scientific) in the attenuated total reflection mode. The *in-operando* tests were performed on the same FTIR spectrometer with an in situ electrochemical cell (PIKE Technologies). The ss-NMR spectra were collected on an NMR spectrometer (Avance III, Bruker). $N_2$ physisorption isotherms were recorded on a gas adsorption analyzer (iQ2, Quantachrome) at 77 K. XRD patterns were collected on an XRD diffractometer (X'Pert, PANalytical) with a Cu Kα X-ray source. Electron microscope images were taken on an SEM (Ultra Plus, Zeiss) and a TEM (JEM-2200, JOEL) under an acceleration voltage of 80 kV. DRS measurement was tested on a UV spectrometer (UV-3600, Shimadzu) using $BaSO_4$ for background correction. XPS and UPS spectra were collected on an XPS spectrometer (K-Alpha+, Thermo Scientific) equipped with an Al-Kα (1486.3 eV) and a He discharge lamp (21.2 eV). All binding energies were corrected with graphite. Mott-Schottky plots were developed by using the capacitance obtained from EIS measurement performed at various potentials in an Ar saturated 0.1 M KOH electrolyte.

*Electrocatalyst performance tests:* The electrochemical performance of electrocatalysts was evaluated using an electrochemical workstation (760E, CHI) using the three-electrode configuration in 0.1 M KOH electrolyte at 25 °C. A Pt mesh and a pre-calibrated Hg/HgO electrode (0.1 M NaOH) served as the counter and the reference electrodes, respectively. All potentials reported were corrected to RHE by adding 0.165 + 0.0591×pH. Electrocatalysts were dispersed in water/isopropanol (v/v:1/9) solution by bath sonication to reach a concentration of 5 mg mL$^{-1}$. The working electrode was prepared by loading 0.2 mg cm$^{-2}$ of electrocatalysts on a pre-polished RRDE (E6R2, Pine Instrument, GC disk OD =5.5 mm, Pt ring ID=6.5 mm and OD=8.5 mm) with a calibrated collection efficiency (*N*) is 0.379. The electrocatalyst was pre-cycled by 20 cyclic voltammetry scans between 0.1~1 $V_{RHE}$ for ORR or 1.2~1.8 $V_{RHE}$ for OER in Ar saturated 0.1 M KOH electrode. Afterwards, LSV curves were obtained in $O_2$ saturated 0.1 M KOH electrolyte at a scan rate of 2 mV s$^{-1}$ without *i*R-compensation.

For ORR tests, Commercial Pt/C (20 wt.%, on Vulcan-XC-72R, Sigma) was tested under identical loading as the references. Selectivity toward $H_2O_2$ formation ($H_2O_2\%$) and the ORR electron transfer number ($n$) were calculated from experimental data obtained using RRDEs in $O_2$ saturated electrolytes by equation (1) and (2), respectively:

$$H_2O_2\% = 200 \times \frac{i_{ring}/N}{i_{disk}+i_{ring}/N} \tag{1}$$

$$n = 4 \times \frac{i_{disk}}{i_{disk}+i_{ring}/N} \tag{2}$$

where $i_{ring}$ and $i_{disk}$ are the currents obtained from the Pt ring and GC disk, respectively. $N$ is the calibrated collection efficiency. The electron transfer number ($n$) and $j_K$ were also calculated using the K-L equation:

$$1/j = 1/j_L + 1/j_K = 1/B\omega^{1/2} + 1/j_K \tag{3}$$

$$B = 0.62nFC_0(D_0)^{2/3}v^{-1/6} \tag{4}$$

$$j_K = nFkC_0 \tag{5}$$

where $j$ is the measured current density, $j_k$ and $j_L$ are the kinetic- and diffusion-limiting current densities, $\omega$ is the angular velocity, $n$ is the transferred electron number, $F$ is the Faraday constant (96485 C mol$^{-1}$), $C_0$ is the saturated concentration of $O_2$ in 0.1 M KOH at room temperature, $D_0$ is diffusion coefficient of oxygen, $v$ is the kinematic viscosity of the electrolyte at room temperature, and $k$ is the electron-transfer rate constant.

For OER tests, the Faradaic efficiency was assessed on a RRDE in Ar saturated 0.1 M KOH electrolytes. The Pt ring was biased at 0.4 $V_{RHE}$, and the GC disk was biased at 1.5 $V_{RHE}$. The $H_2O_2$ selectivity was calculated by equation (1) to determine the catalyst efficiency towards the 4-electron OER. The OER TOF of electrocatalysts was calculated by assuming all S-C regions in COF shells are active sites, using equation (6) shown below:

$$TOF = \frac{i}{4 \times 96485 \times n} \tag{6}$$

where $i$ is the current, and $n$ is the number of active sites, which is calculated by equation (7):

$$n = \frac{2 \times m \times w}{M_w} \quad (7)$$

where $m$ is the mass of electrocatalysts loaded on the electrode, $w$ is the mass fraction of COF in electrocatalysts, $M_w$ is the molecular weight of a single unit of TAPTt-COF (~370 g mol$^{-1}$), containing two identical C-S regions. The electrodes for the water electrolyzer were prepared by coating 0.2 mg cm$^{-2}$ electrocatalysts on carbon clothes with a geometric dimension of 1 cm$^2$.

*Computational method:* Density functional theory (DFT) calculations in this study were performed using the VASP code. Electron correlation was calculated using the generalized gradient approximation method with the functional developed by Perdew, Burke, and Ernzerhof. Core electrons were considered using the projector augmented wave method.[28] The valence electrons were described by expanding the Kohn-Sham wave functions in a plane-wave basis set, with a kinetic cutoff of 400 eV.[29]

Convergence was established when the forces of each atom were lower than 0.05 eV Å$^{-1}$. A gamma-point sampling was used for all calculations. The reaction free energies of ORR were calculated using the computational hydrogen electrode method.[23]

*Zn-air battery tests:* Air electrodes were prepared by depositing electrocatalysts on carbon cloth gas-diffusion layers with a mass loading of 0.5 mg cm$^{-2}$. Reference air electrodes were prepared by depositing commercial Pt/C and IrO$_2$ catalysts (with a 1/1 molar ratio between Pt/Ir) at the same mass loading. A 6 M KOH solution with 0.2 M ZnCl$_2$ was used as the electrolyte. The total volume of electrolytes in every cell is about 20 mL. A piece of Zn foil (0.2 mm in thickness, 99.9%, Sigma) was used as the Zn electrode. The battery performance was evaluated in ambient air at ~25 °C using a battery tester (CT2001, Land).

**Supporting Information**

Supporting Information is available after the main text.

**Conflict of Interest**

The authors declare no conflict of interest.

**Keywords**

covalent organic framework; carbon nanotube; van der Waals heterostructure; oxygen redox reaction; zinc-air battery

**Acknowledgments**

We thank funding support from the Australian Research Council under the Future Fellowships scheme (FT160100107) and Discovery Programme (DP180102210), the support from GDAS' Special Project of Science and Technology Development (2019GDASYL-0104005 and 2020GDASYL-20200402001). H. Li and G. Henkelman acknowledge the Welch Foundation (F-1841) and the Texas Advanced Computing Center for computational resources.

**Reference**

[1] X. Zhao, P. Pachfule, S. Li, T. Langenhahn, M. Ye, C. Schlesiger, S. Praetz, J. Schmidt, A. Thomas, J. Am. Chem. Soc. 2019, 141, 6623; J. Guo, C. Y. Lin, Z. Xia, Z. Xiang, Angew. Chem. Int. Ed. 2018, 57, 12567; S. Yang, Y. Yu, M. Dou, Z. Zhang, L. Dai, F. Wang, Angew. Chem. Int. Ed. 2019, 58, 14724; P. Peng, L. Shi, F. Huo, C. Mi, X. Wu, S. Zhang, Z. Xiang, Sci. Adv. 2019, 5, eaaw2322; B. Q. Li, S. Y. Zhang, B. Wang, Z. J. Xia, C. Tang, Q. Zhang, Energ Environ Sci 2018, 11, 1723; P. Peng, Z. Zhou, J. Guo, Z. Xiang, ACS Energy Lett. 2017, 2, 1308.

[2] X. Feng, X. Ding, D. Jiang, Chem. Soc. Rev. 2012, 41, 6010; S. Y. Ding, W. Wang, Chem. Soc. Rev. 2013, 42, 548.

[3] X. Cui, S. Lei, A. C. Wang, L. Gao, Q. Zhang, Y. Yang, Z. Lin, Nano Energy 2020, 70, 104525.

[4] D. J. Yang, L. J. Zhang, X. C. Yan, X. D. Yao, Small Methods 2017, 1; X. Chen, Z. Zhou, H. E. Karahan, Q. Shao, L. Wei, Y. Chen, Small 2018, 14, e1801929; Z. F. Huang, J. Wang, Y. C. Peng, C. Y. Jung, A. Fisher, X. Wang, Adv Energy Mater 2017, 7, 1700544; J. Fu, Z. P. Cano, M. G. Park, A. Yu, M. Fowler, Z. Chen, Adv. Mater. 2017, 29; Z. F. Pan, L. An, T. S. Zhao, Z. K. Tang, Prog Energ Combust 2018, 66, 141; H. Du, C. X. Zhao, J. Lin, J. Guo, B. Wang, Z. Hu, Q. Shao, D. Pan, E. K. Wujcik, Z. Guo, Chem. Rec. 2018, 18, 1365; J. Qi, W. Zhang, R. Cao, Adv Energy Mater 2018, 8; I. Vincent, D. Bessarabov, Renew Sust Energ Rev 2018, 81, 1690.

[5] X. Li, Q. Gao, J. Aneesh, H. S. Xu, Z. X. Chen, W. Tang, C. B. Liu, X. Y. Shi, K. V. Adarsh, Y. X. Lu, K. P. Loh, Chem Mater 2018, 30, 5743; E. Jin, M. Asada, Q. Xu, S. Dalapati, M. A. Addicoat, M. A. Brady, H. Xu, T. Nakamura, T. Heine, Q. Chen, Science 2017, 357, 673.

[6] J. Liang, Y. Zheng, J. Chen, J. Liu, D. Hulicova-Jurcakova, M. Jaroniec, S. Z. Qiao, Angew. Chem. Int. Ed. 2012, 51, 3892; Y. Zheng, Y. Jiao, Y. Zhu, L. H. Li, Y. Han, Y. Chen, A. Du, M. Jaroniec, S. Z. Qiao, Nat. Commun. 2014, 5, 3783; S. Chen, J. Duan, M. Jaroniec, S.-Z. Qiao, Adv. Mater. 2014, 26, 2925.

[7] G. L. Tian, M. Q. Zhao, D. Yu, X. Y. Kong, J. Q. Huang, Q. Zhang, F. Wei, Small 2014, 10, 2251.

[8] W. H. Niu, K. Marcus, L. Zhou, Z. Li, L. Shi, K. Liang, Y. Yang, Acs Catal 2018, 8, 1926.

[9] R. Gao, Q. Dai, F. Du, D. Yan, L. Dai, J. Am. Chem. Soc. 2019, 141, 11658.

[10] S. Wang, D. Yu, L. Dai, J. Am. Chem. Soc. 2011, 133, 5182; Y. Zhang, X. L. Fan, J. H. Jian, D. S. Yu, Z. S. Zhang, L. M. Dai, Energ Environ Sci 2017, 10, 2312; Q. Hu, G. Li, X. Liu, B. Zhu, X. Chai, Q. Zhang, J. Liu, C. He, Angew. Chem. Int. Ed. 2019, 58, 4318.

[11] D. Jariwala, T. J. Marks, M. C. Hersam, Nat. Mater. 2017, 16, 170; Y. Liu, N. O. Weiss, X. D. Duan, H. C. Cheng, Y. Huang, X. F. Duan, Nat Rev Mater 2016, 1, 16042.

[12] R. Xiang, T. Inoue, Y. Zheng, A. Kumamoto, Y. Qian, Y. Sato, M. Liu, D. Tang, D. Gokhale, J. Guo, K. Hisama, S. Yotsumoto, T. Ogamoto, H. Arai, Y. Kobayashi, H. Zhang, B. Hou, A.

Anisimov, M. Maruyama, Y. Miyata, S. Okada, S. Chiashi, Y. Li, J. Kong, E. I. Kauppinen, Y. Ikuhara, K. Suenaga, S. Maruyama, Science 2020, 367, 537.

[13] A. M. El-Sawy, I. M. Mosa, D. Su, C. J. Guild, S. Khalid, R. Joesten, J. F. Rusling, S. L. Suib, Adv Energy Mater 2016, 6, 1501966.

[14] N. Nakashima, Y. Tomonari, H. Murakami, Chem. Lett. 2002, 31, 638.

[15] X. Lu, W.-L. Yim, B. H. R. Suryanto, C. Zhao, J. Am. Chem. Soc. 2015.

[16] J. Suntivich, K. J. May, H. A. Gasteiger, J. B. Goodenough, Y. Shao-Horn, Science 2011, 334, 1383.

[17] Q. Li, T. He, Y.-Q. Zhang, H. Wu, J. Liu, Y. Qi, Y. Lei, H. Chen, Z. Sun, C. Peng, ACS Sustainable Chemistry & Engineering 2019, 7, 17039; Q. Liu, Y. B. Wang, L. M. Dai, J. N. Yao, Adv. Mater. 2016, 28, 3000; T. Y. Ma, J. Ran, S. Dai, M. Jaroniec, S. Z. Qiao, Angew. Chem. Int. Ed. 2014, 54, 1; S. Liu, H. Zhang, Q. Zhao, X. Zhang, R. Liu, X. Ge, G. Wang, H. Zhao, W. Cai, Carbon 2016, 106, 74; Y. Qian, Z. Hu, X. Ge, S. Yang, Y. Peng, Z. Kang, Z. Liu, J. Y. Lee, D. Zhao, Carbon 2017, 111, 641; Z. Lu, J. Wang, S. Huang, Y. Hou, Y. Li, Y. Zhao, S. Mu, J. Zhang, Y. Zhao, Nano Energy 2017, 42, 334; K. Qu, Y. Zheng, S. Dai, S. Z. Qiao, Nano Energy 2016, 19, 373; N. Jia, Q. Weng, Y. Shi, X. Shi, X. Chen, P. Chen, Z. An, Y. Chen, Nano Res. 2018, 11, 1905; J. Zhang, Z. Zhao, Z. Xia, L. Dai, Nat. Nanotechnol. 2015, 10, 444; H. B. Yang, J. Miao, S.-F. Hung, J. Chen, H. B. Tao, X. Wang, L. Zhang, R. Chen, J. Gao, H. M. Chen, L. Dai, B. Liu, Sci. Adv. 2016, 2; X. Lin, P. Peng, J. Guo, Z. Xiang, Chem. Eng. J. 2019, 358, 427; Y. Guo, S. Yao, L. Gao, A. Chen, M. Jiao, H. Cui, Z. Zhou, J Mater Chem A 2020, 8, 4386; H.-F. Wang, C. Tang, Q. Zhang, Catal. Today 2018, 301, 25; X. Xiao, X. Li, Z. Wang, G. Yan, H. Guo, Q. Hu, L. Li, Y. Liu, J. Wang, Appl. Catal. B 2020, 265, 118603.

[18] Z. Wang, B. Yang, Y. Wang, Y. Zhao, X. M. Cao, P. Hu, Phys. Chem. Chem. Phys. 2013, 15, 9498.

[19] S. M. Sze, K. K. Ng, in *Physics of Semiconductor Devices*, John Wiley & Sons, 2006, 5.

[20] F. Cardon, W. Gomes, J. Phys. D Appl. Phys. 1978, 11, L63.

[21] J. Hermann, D. Alfè, A. Tkatchenko, Nat. Commun. 2017, 8, 14052.

[22] J. Y. Cheon, J. H. Kim, J. H. Kim, K. C. Goddeti, J. Y. Park, S. H. Joo, J. Am. Chem. Soc. 2014, 136, 8875.

[23] J. K. Nørskov, J. Rossmeisl, A. Logadottir, L. Lindqvist, J. R. Kitchin, T. Bligaard, H. Jónsson, J. Phys. Chem. B 2004, 108, 17886.

[24] J. A. Gauthier, C. F. Dickens, L. D. Chen, A. D. Doyle, J. K. Nørskov, J. Phys. Chem. C 2017, 121, 11455.

[25] V. Vacque, B. Sombret, J. P. Huvenne, P. Legrand, S. Suc, Spectrochim. Acta A 1997, 53, 55.

[26] T. Noguchi, M. Nojiri, K.-i. Takei, M. Odaka, N. Kamiya, Biochemistry 2003, 42, 11642.

[27] G. Beamson, D. Briggs, *High Resolution XPS of Organic Polymers: The Scienta ESCA300 Database*, Wiley, 1992.

[28] B. Hammer, L. B. Hansen, J. K. Nørskov, Phys. Rev. B 1999, 59, 7413; J. P. Perdew, K. Burke, M. Ernzerhof, Phys. Rev. Lett. 1996, 77, 3865.

[29] W. Kohn, L. J. Sham, Phys. Rev. 1965, 140, A1133.

**TOC**

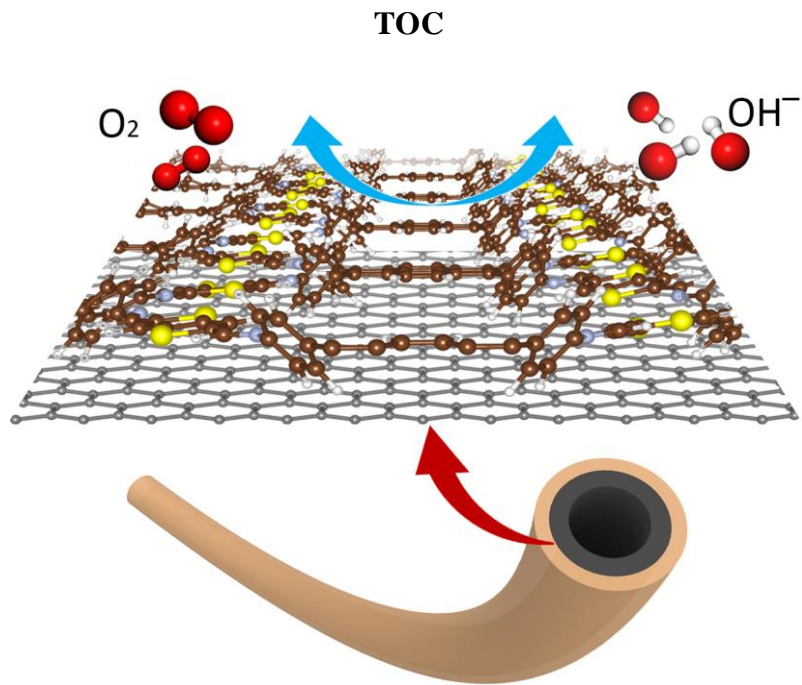

A coaxial 1D van der Waals heterostructure comprises a carbon nanotube (CNT) core and a covalent organic framework (COF) shell. This 1D coaxial structure promotes electronic interactions between CNT and COF. It suppresses the stacking of COF, resulting in significantly improved catalytic activity for oxygen redox reactions at the S-C active centers in the pre-designed COF, which enables high-performance rechargeable zinc-air batteries.

**Supporting Information**

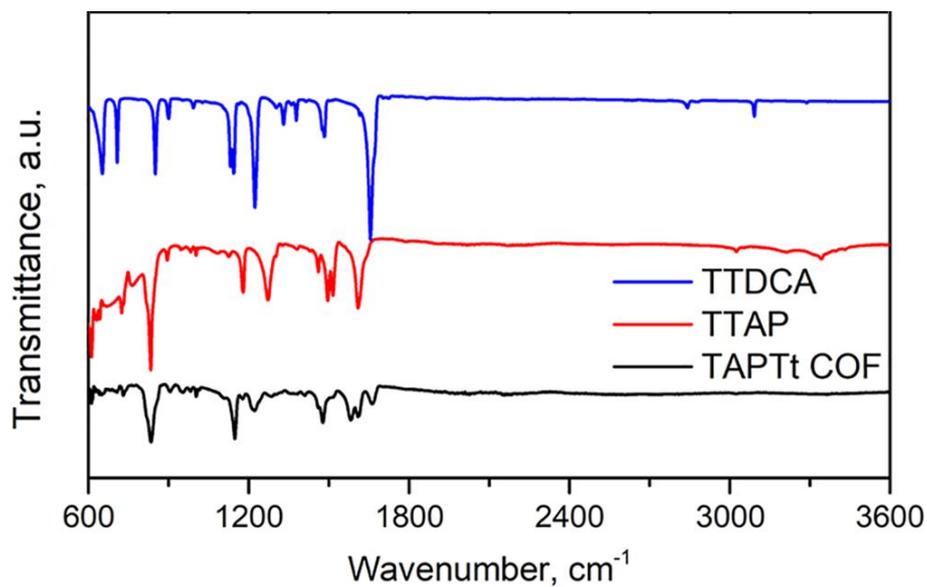

**Figure S1.** FTIR spectra of precursor molecules (TTDCA and TTAP) and TAPTt-COF. The peak at 1658 cm$^{-1}$ can be assigned to imine (-C=N-) bonds in the framework. The peaks between 600~800 cm$^{-1}$ and the peak at 834 cm$^{-1}$ are originated from C–S, and C–H vibrations in thienothiophene, respectively.

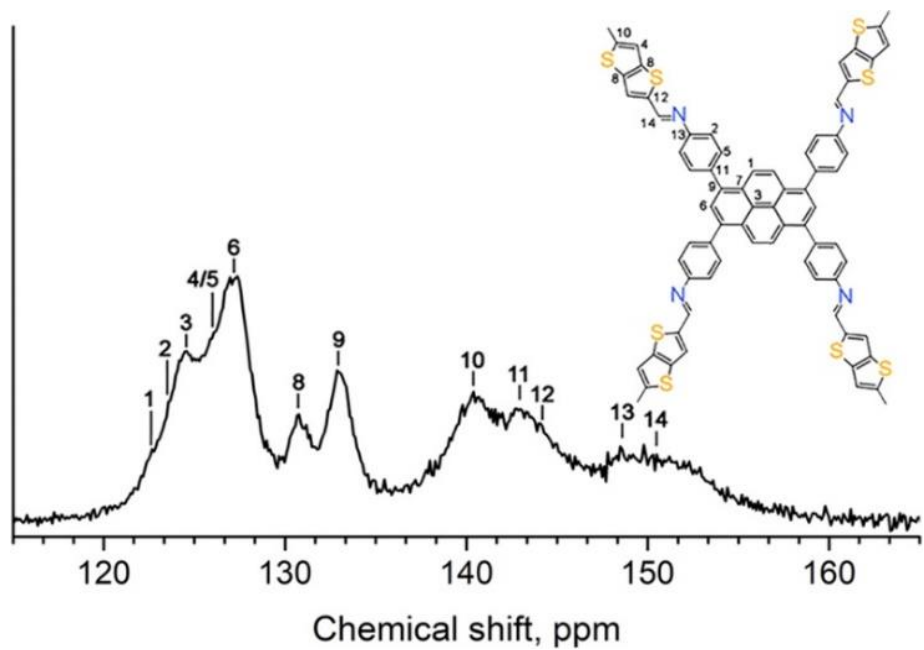

**Figure S2.** Solid-state NMR spectrum of TAPTt-COF.

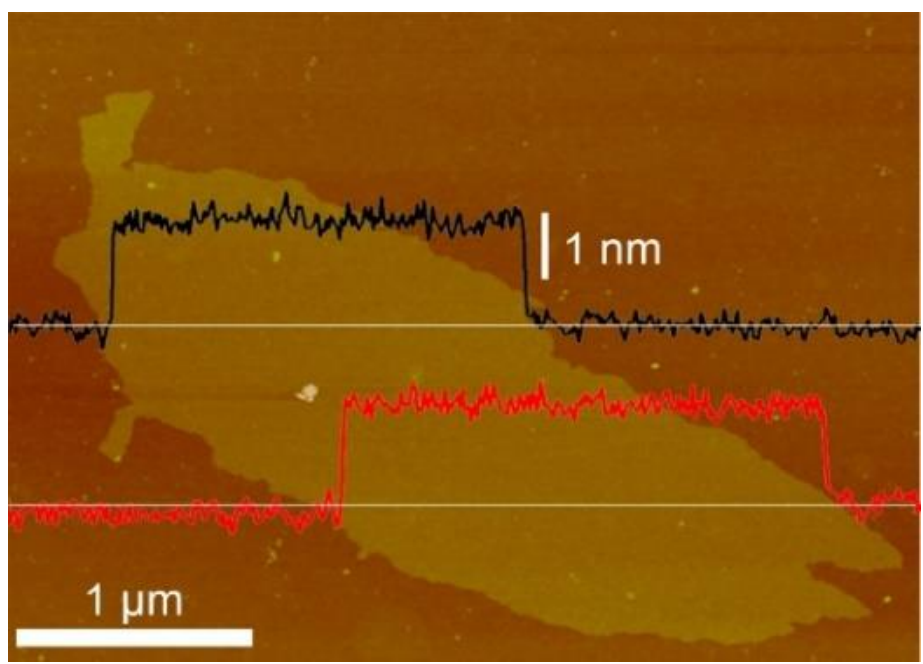

**Figure S3.** An AFM image and the corresponding height profiles of an individual TAPTt-COF nanosheet.

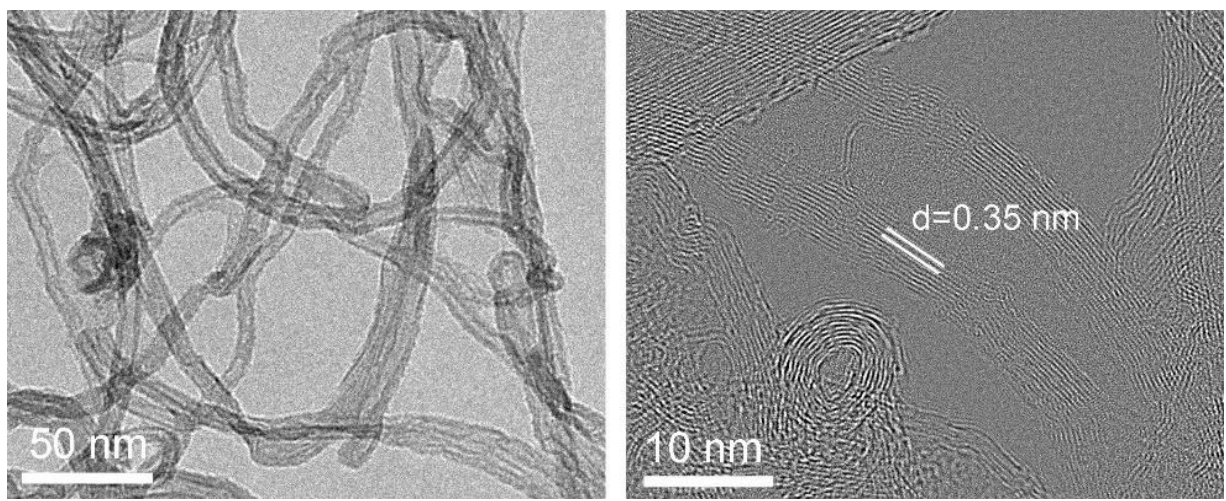

**Figure S4.** TEM images of pristine CNTs used in this study.

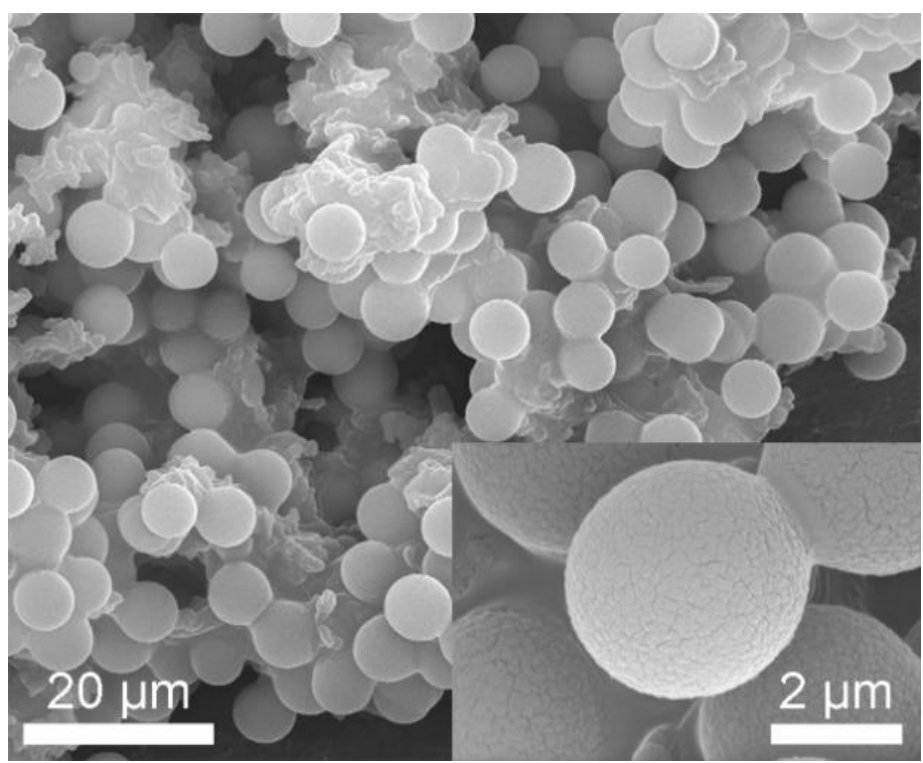

**Figure S5.** SEM images of aggregated TAPTt-COF without the presence of CNTs.

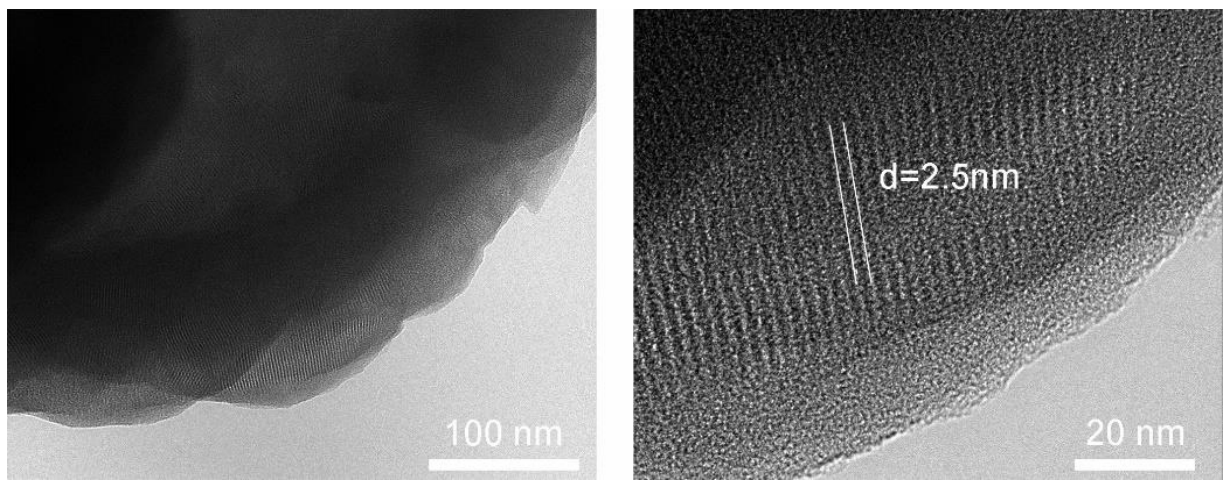

**Figure S6.** TEM images of aggregated TAPTt-COF without the presence of CNTs.

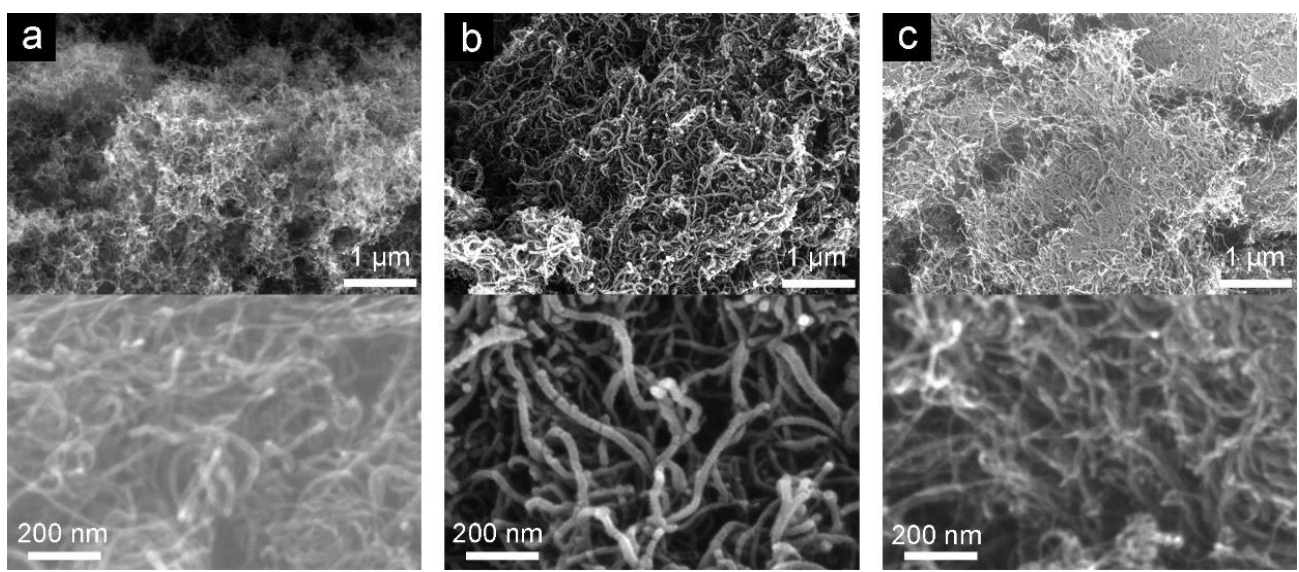

**Figure S7.** SEM images of (a) CC-4, (b) CC-3 and (c) CC-2 vdWHs.

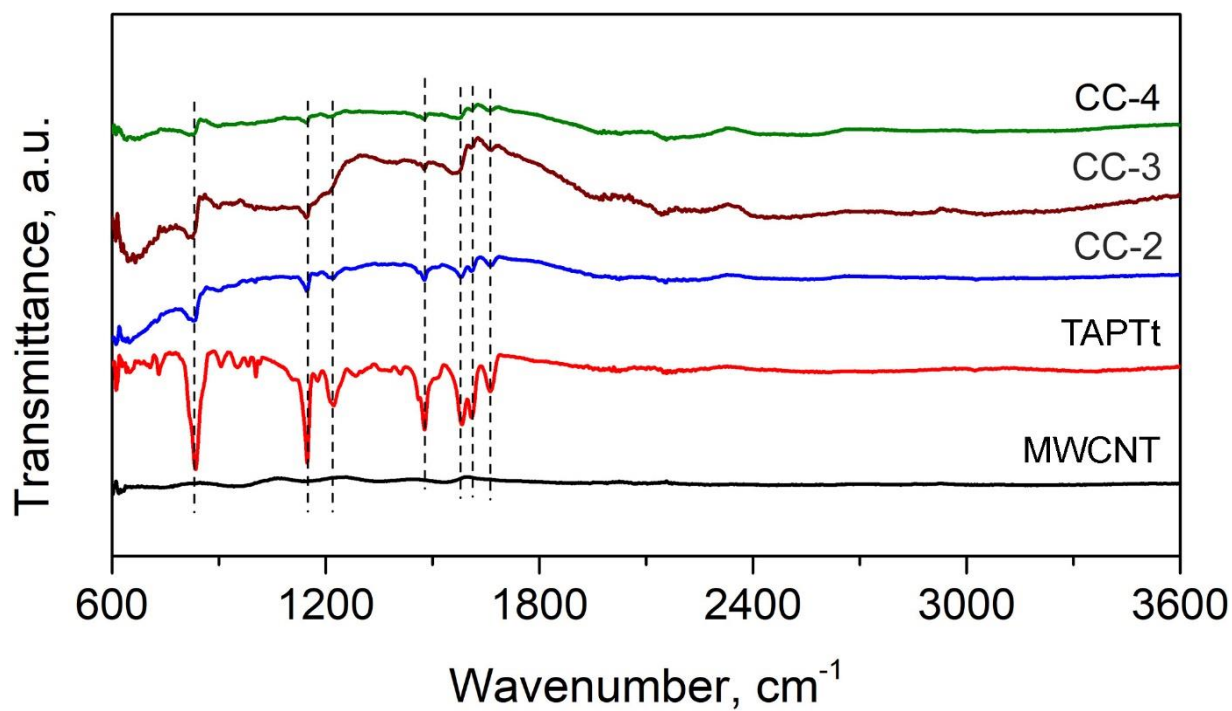

**Figure S8.** FTIR spectra of TAPTt-COF, MWCNT, and CC-X vdWHs.

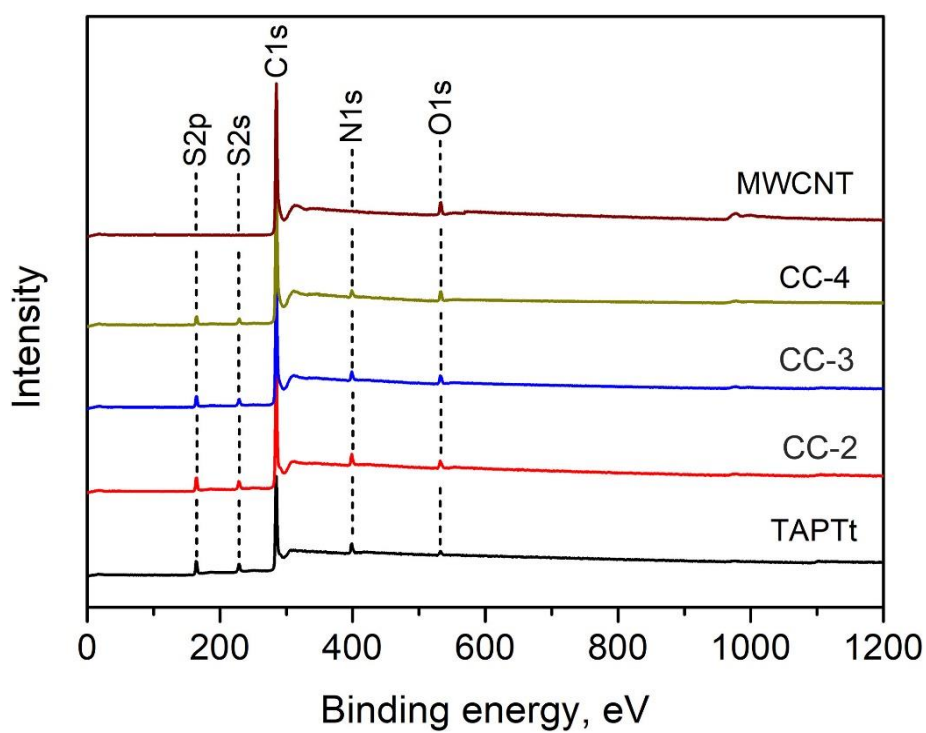

**Figure S9.** XPS survey scans of pristine TAPTt-COF, MWCNT, and CC-X vdWHs.

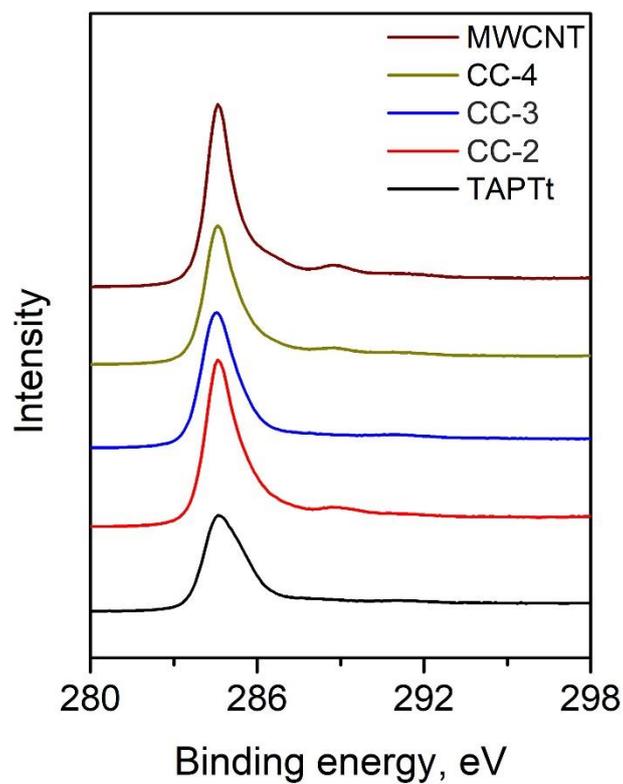

**Figure S10.** High-resolution XPS C1s spectra of pristine TAPTt-COF, MWCNT, and CC-x VdWHs.

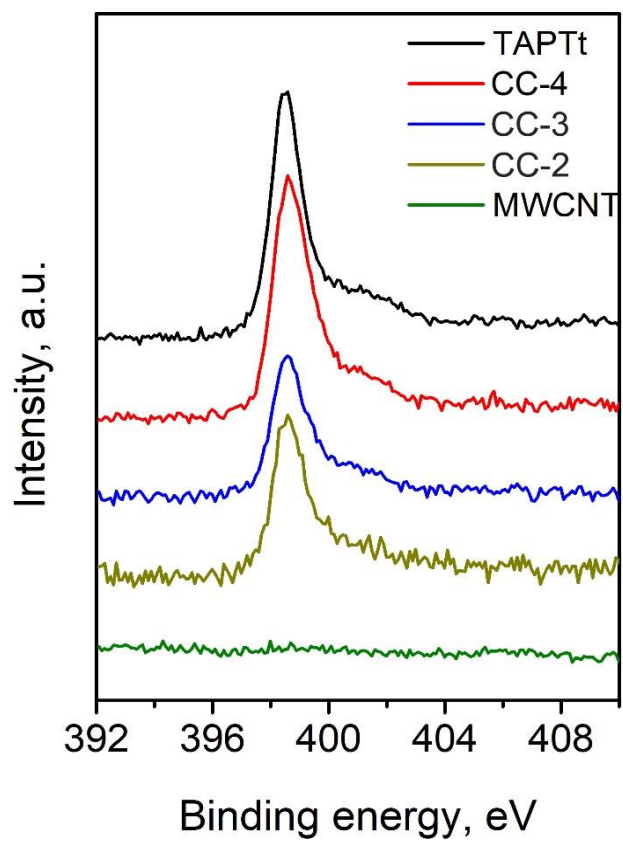

**Figure S11.** High-resolution XPS N1s spectra of pristine TAPTt-COF, MWCNT, and CC-x VdWHs.

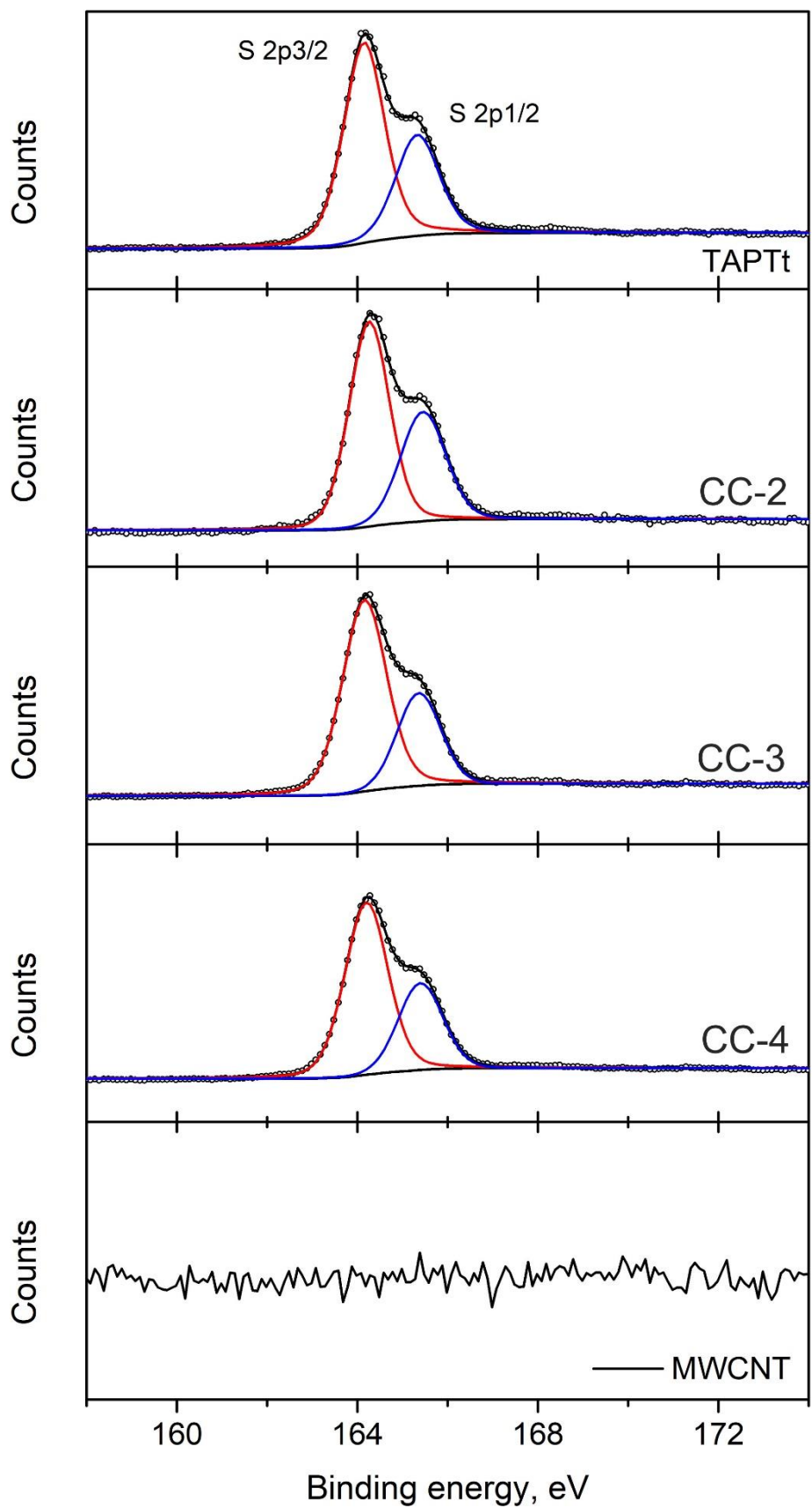

**Figure S12.** High-resolution XPS S2p spectra of pristine TAPTt-COF, MWCNT, and CC-x VdWHs.

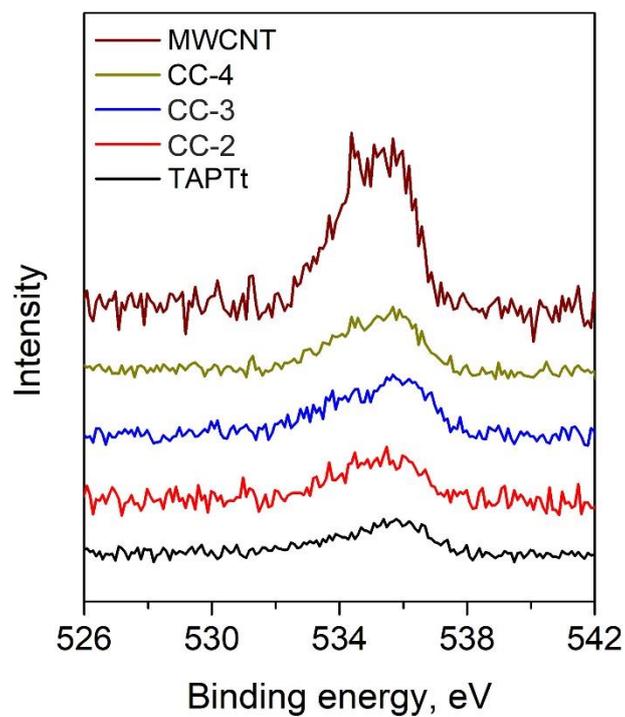

**Figure S13.** High-resolution XPS O spectra of pristine TAPTt-COF, MWCNT, and CC-x VdWHs.

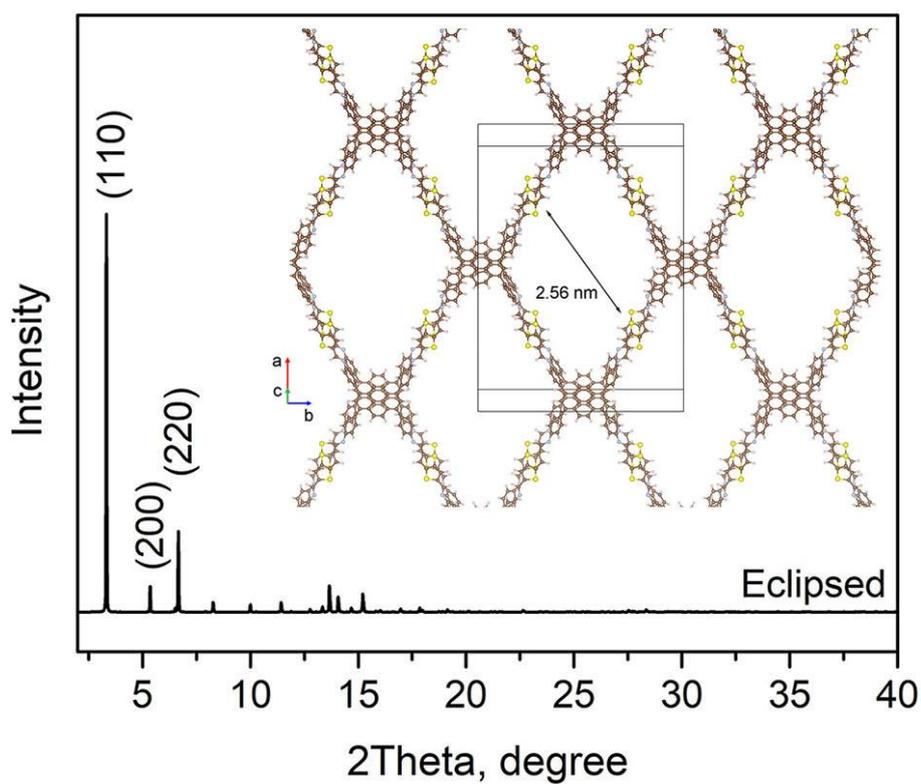

**Figure S14.** The eclipsed model of TAPTt COF and its corresponding simulated XRD pattern.

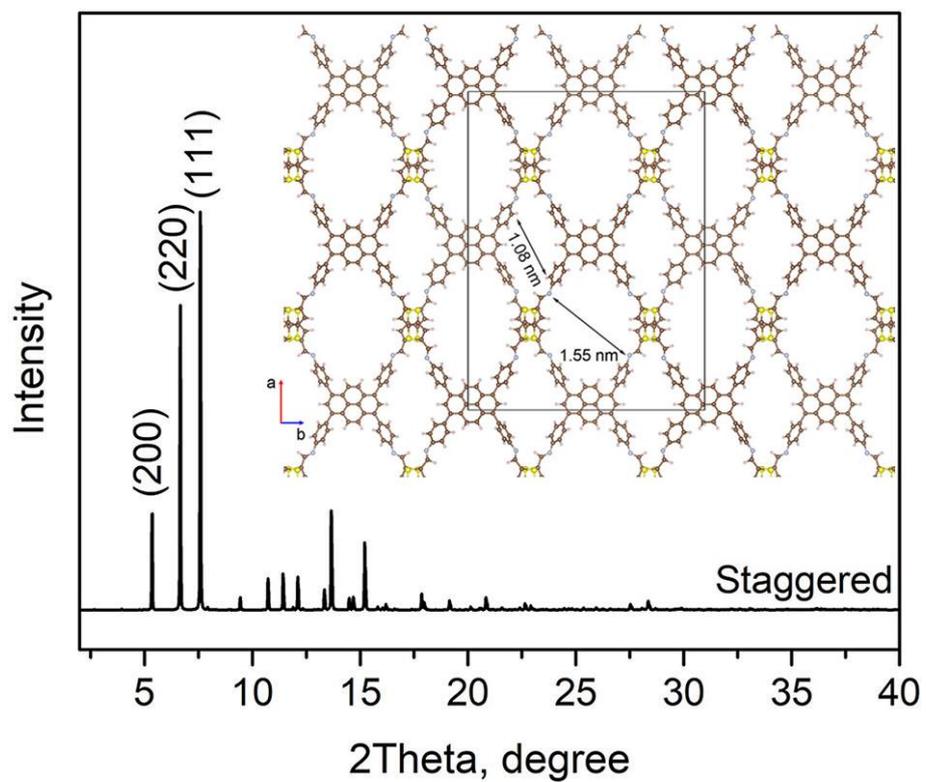

**Figure S15.** The staggered model of TAPTt COF and its corresponding simulated XRD pattern.

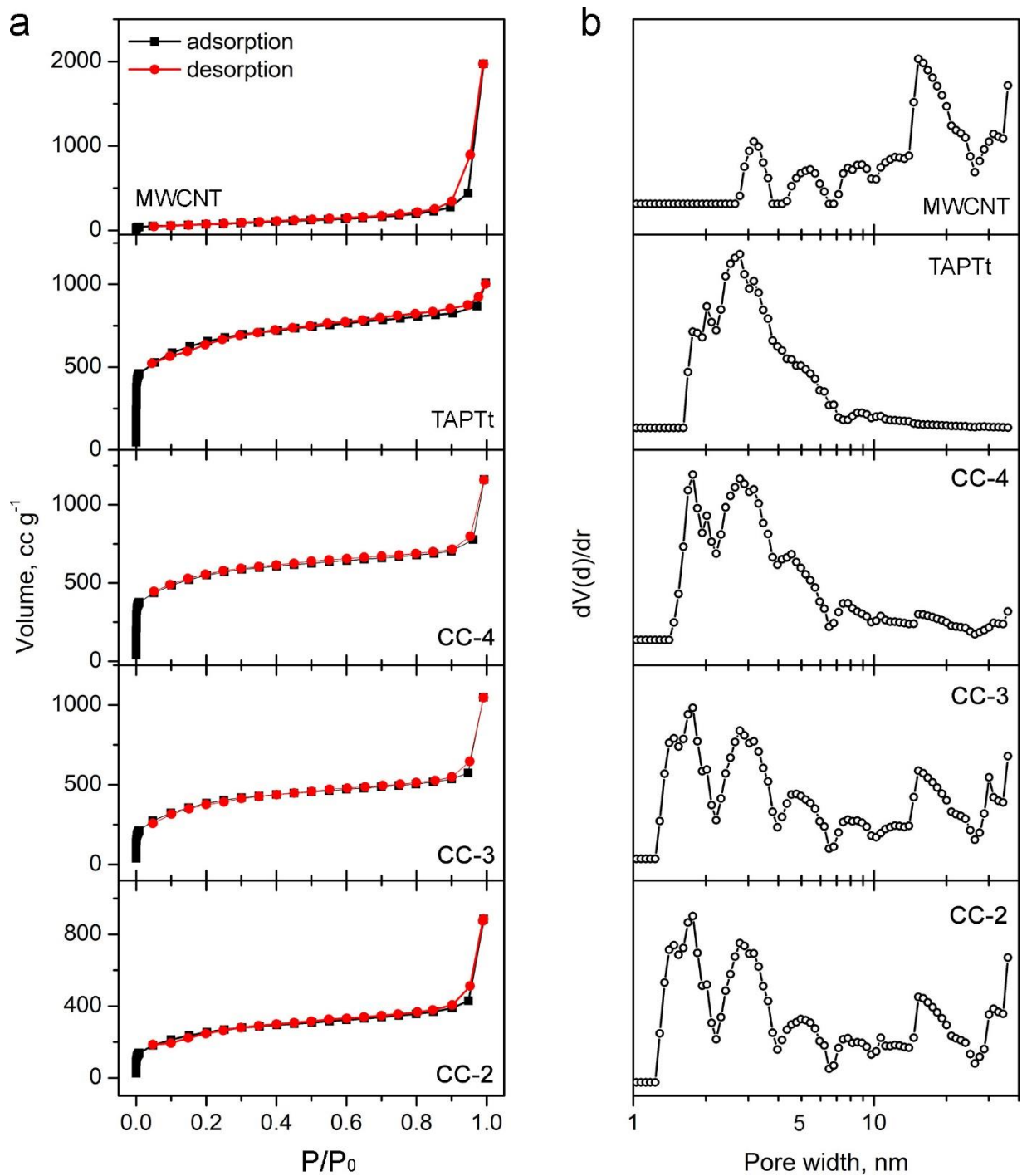

**Figure S16.** (a) N$_2$ physisorption isotherms and (b) pore-size distributions of TAPTt-COF, MWCNT, and CC-x VdWHs, determined by the DFT method.

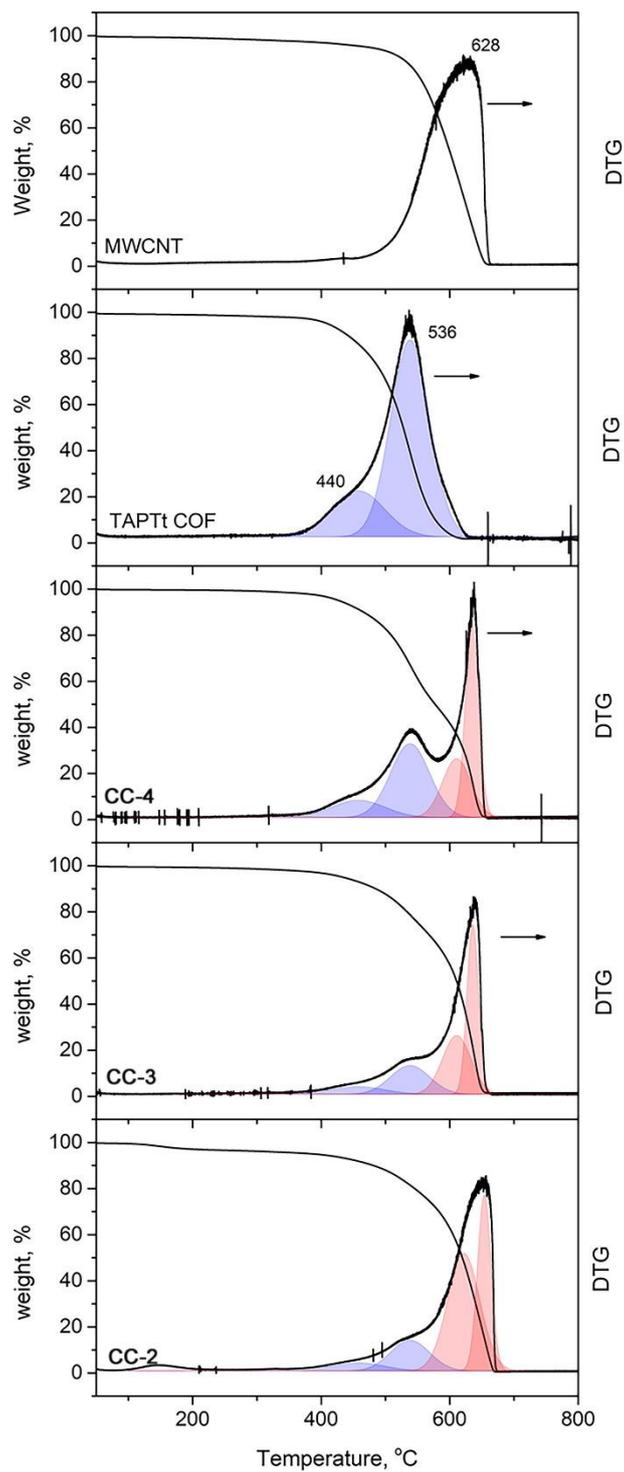

**Figure S17.** TGA and DTG profiles of TAPTt-COF, MWCNT, and CC-x VdWHs.

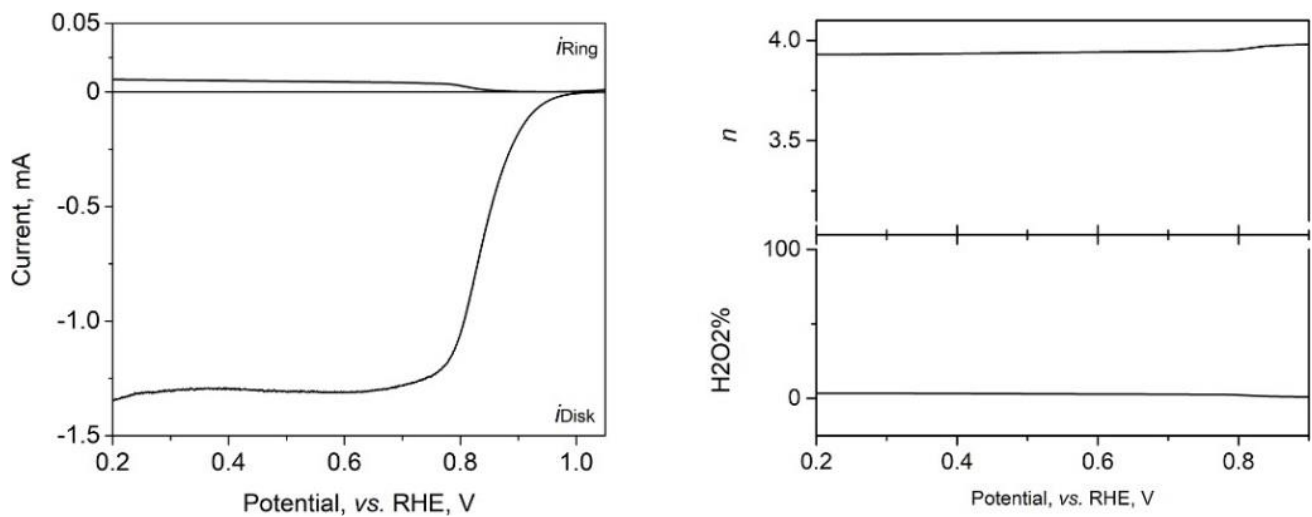

**Figure S18.** ORR RRDE-LSV curves of 20 wt.% Pt/C and its calculated *n* and the selectivity to $H_2O_2$.

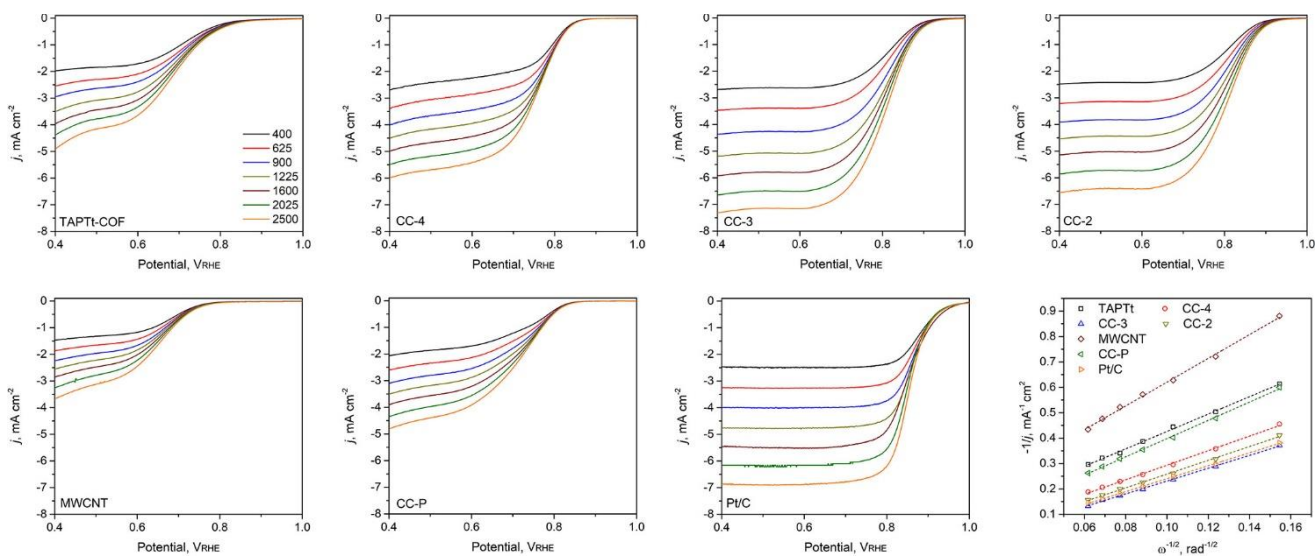

**Figure S19.** ORR RDE LSV plots of different samples and their corresponding K-L plot at 0.7 $V_{RHE}$.

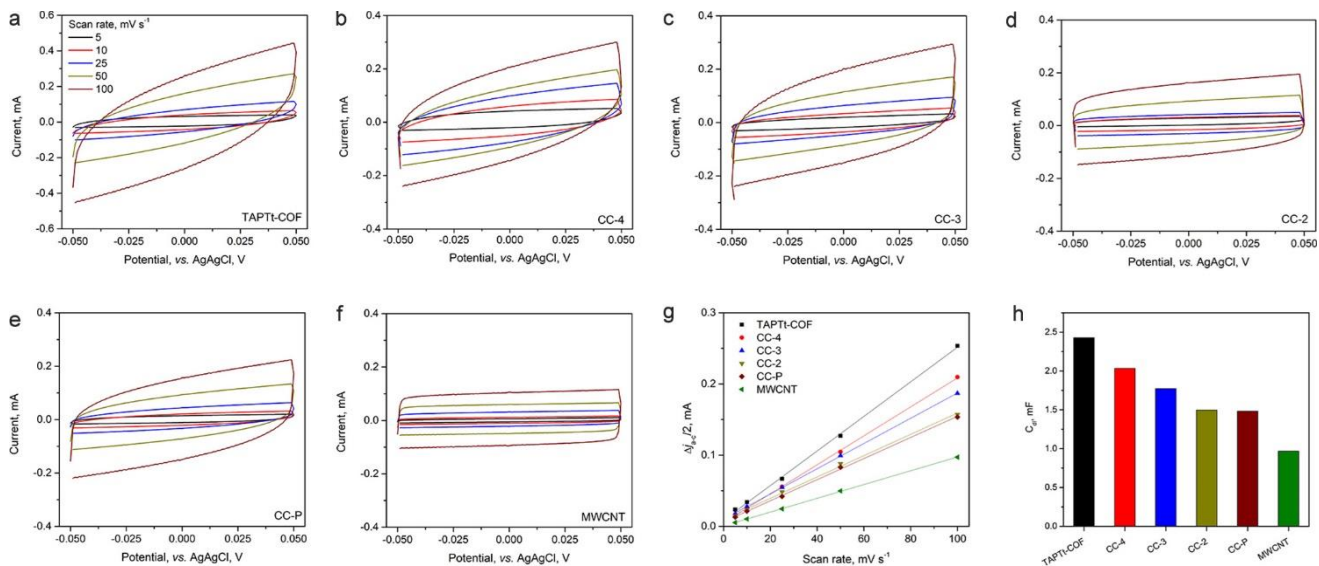

**Figure S20.** CV curves of (a) TAPTt-COF, (b) CC-4, (c) CC-3, (d) CC-2, (e) CC-P and (f) CNTs obtained at different scan rates from 5 to 100 mV s$^{-1}$. (g) Linear fitting of current differences at 0 V *vs.* a reference Hg/HgO electrode and (h) calculated $C_{dl}$ of different samples.

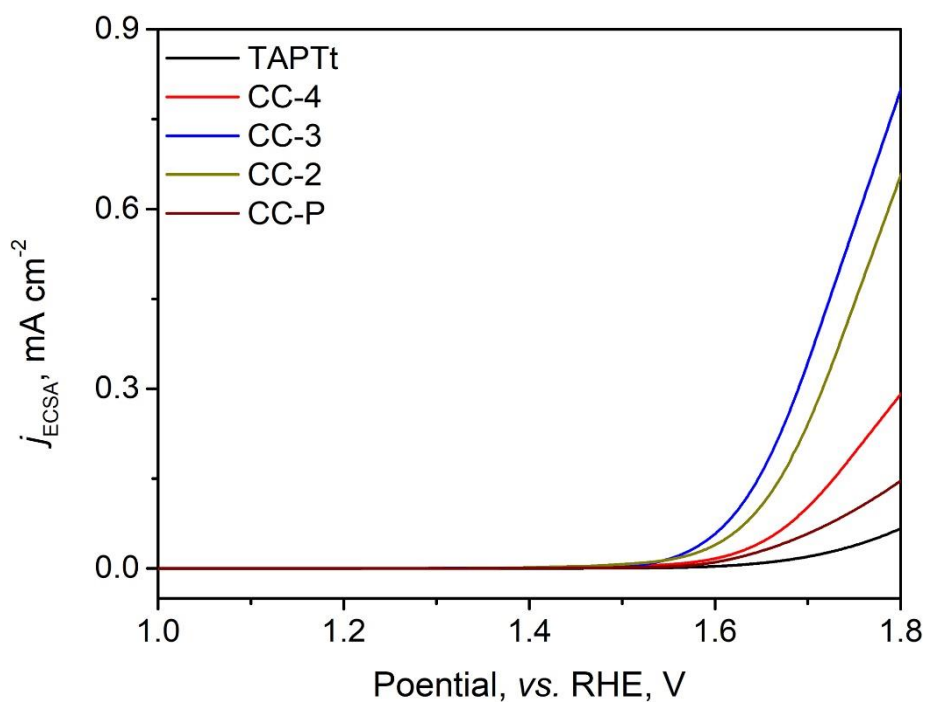

**Figure S21.** ECSA normalized OER LSV curves of TAPTt-COF, CC-x, and CC-P.

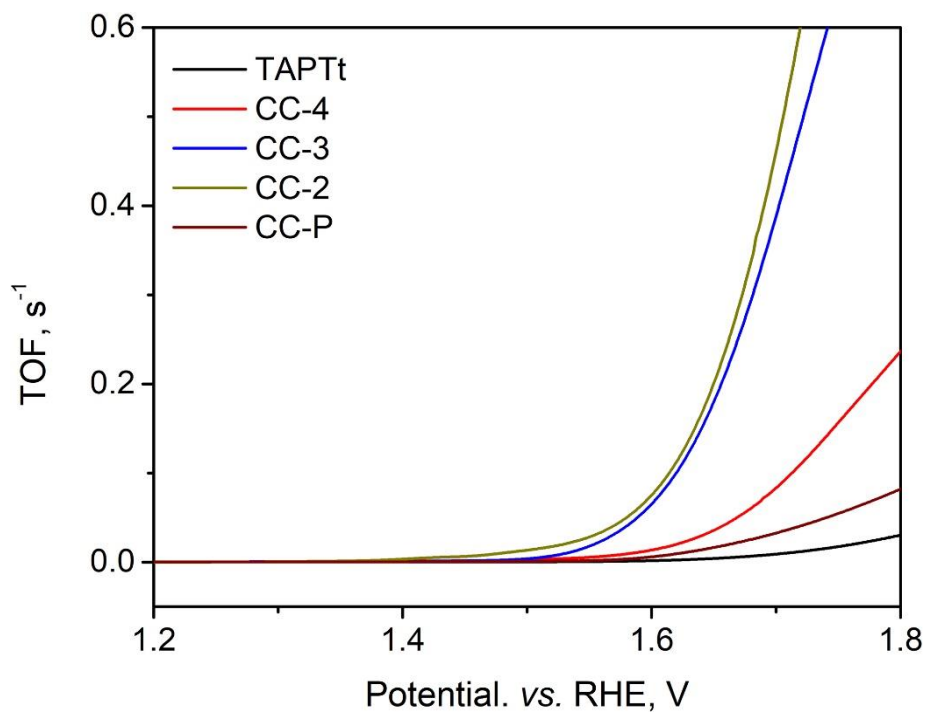

**Figure S22.** OER TOFs of TAPTt-COF, CC-x, and CC-P.

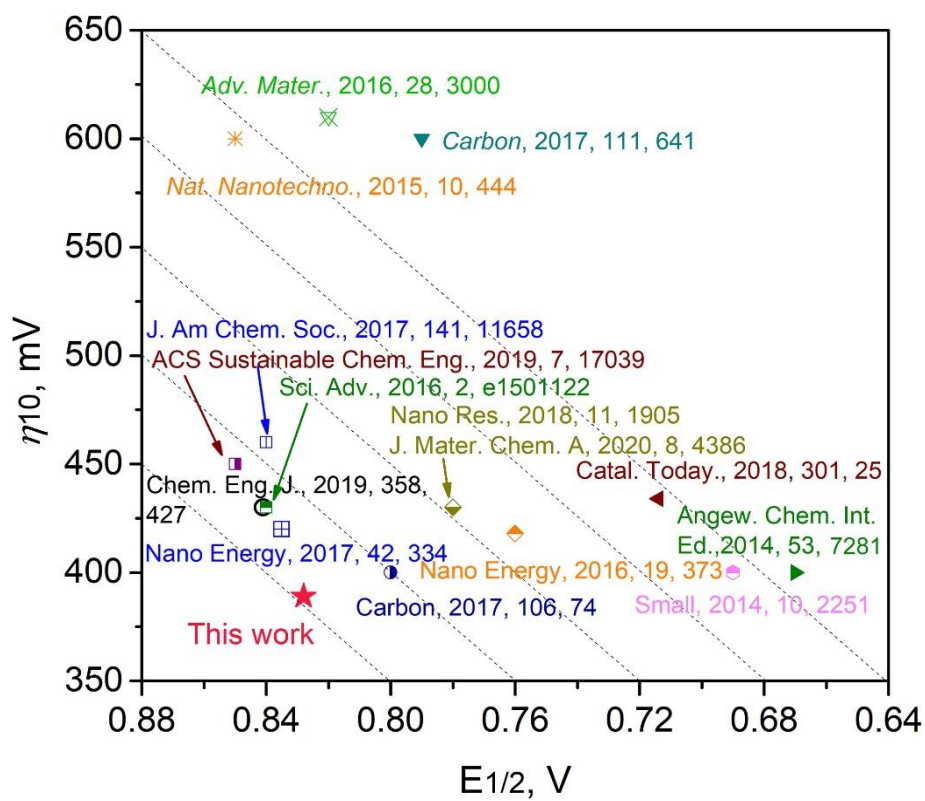

**Figure S23.** Activity atlas of various bifunctional carbon electrocatalyst by plotting their ORR $E_{1/2}$ against OER $\eta_{10}$ values obtained in 0.1 M KOH electrolyte on rotary disk electrodes. [1-20]

**Table S1.** Performance comparison of some recently reported metal-free and carbon-based ORR and OER electrocatalysts in 0.1 M KOH electrolyte.

| Catalyst | ORR performance | | OER performance | | $\Delta E^{e}$, mV | Ref. |
|---|---|---|---|---|---|---|
| | $E_{1/2}$, vs. RHE, V | $n$ | $\eta_{10}$, mV | Tafel slope, mV dec$^{-1}$ | | |
| CC-3 | 0.828 | 3.86 | 389 | 101.5 | 791 | **This work** |
| SNC | 0.78 | 3.92 | 430 | 110 | 880 | 1 |
| N-CNSP | 0.85 | ~4 | 390 $^a$ | - | 770 | 2 |
| HHPC | 0.78 | 3.85 | 350 | - | 800 | 3 |
| BRC$_{AC}$850$_2$ | 0.850 | 3.63 | 450 | - | 830 | 4 |
| NPC-"Cs" | 0.85 | 3.9 | 343 $^d$ | 78 | 723 | 5 |
| RM-COP-PA-900 | 0.841 | 3.87 | 430 | - | 819 | 6 |
| NMGF | 0.714 | 3.6 | 434 | - | 950 | 7 |
| B, N-Carbon | 0.84 | 3.80 | 340 $^a$ | - | 730 | 8 |
| N-CCs | 0.78 $^c$ | ~4 | 430 $^c$ | - | 880 | 9 |
| BNPC-1100 | ~0.79 | 3.6 | ~600 | 201 | 1040 | 10 |
| N-PC@G-0.02 | 0.8 | 3.93 | 400 | 78 | 830 | 11 |
| NB-CN | 0.835 | 3.9 | 420 | - | 815 | 12 |
| C$_{60}$-SWCNT$_{15}$ | 0.84 | - | 460 | 46.7 | 850 | 13 |
| N-GRW | 0.84 | - | 430 | - | 820 | 14 |
| NCNF-1000 | 0.82 | - | 610 | - | 1020 | 15 |
| S,S'-CNT $_{1000\ °C}$ | 0.79 | - | 350 $^a$ | 95 $^a$ | 790 | 16 |
| N,S-CN | 0.76 $^b$ | 3.89 | 418 | 53 | 888 | 17 |
| NMC-1000 | 0.85 | ~4 | ~600 | - | 980 | 18 |
| NGSH | ~0.69 | ~3.6 | 400 | 83 | 940 | 19 |
| P-doped g-C$_3$N$_4$ | 0.67 | - | 400 | - | 960 | 20 |

$^a$· tested in 1 M KOH electrolyte; $^b$· -0.2 V and 0.68 V *vs.* Ag/AgCl (4M KCl, 0.1 M KOH), derived by adding $0.2 + 0.0591*13$; $^c$ tested in 0.1 M NaOH electrolyte; $^d$· tested on nickel foam electrode in 1 M KOH; $^e$· $\Delta E = \eta_{10} - E_{1/2} + 1230$, performance collected in electrolyte other than 0.1 M alkaline solution was not calculated.

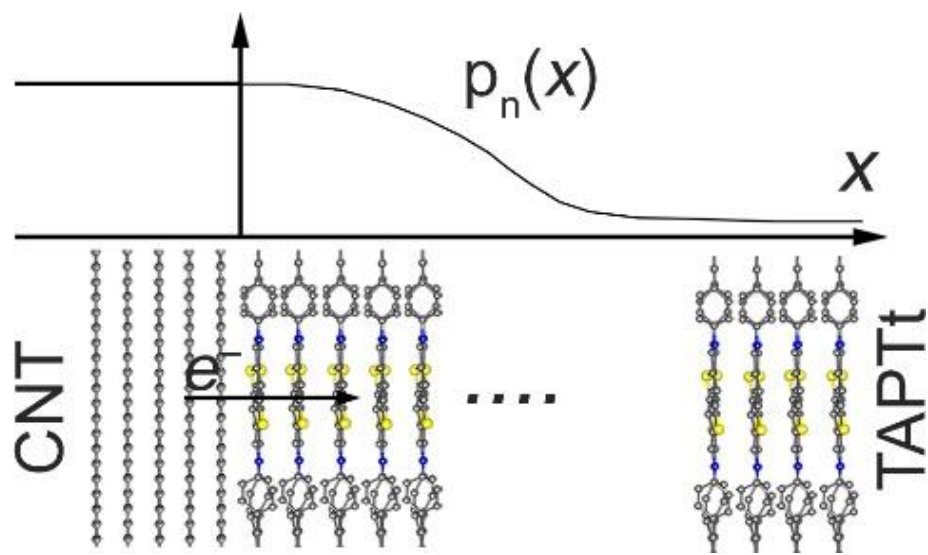

**Figure S24**. The schematic illustration of carrier injection from CNTs to a semi-infinite COF shell at steady-state, where $p_n(x)$ is the carrier density function.[21]

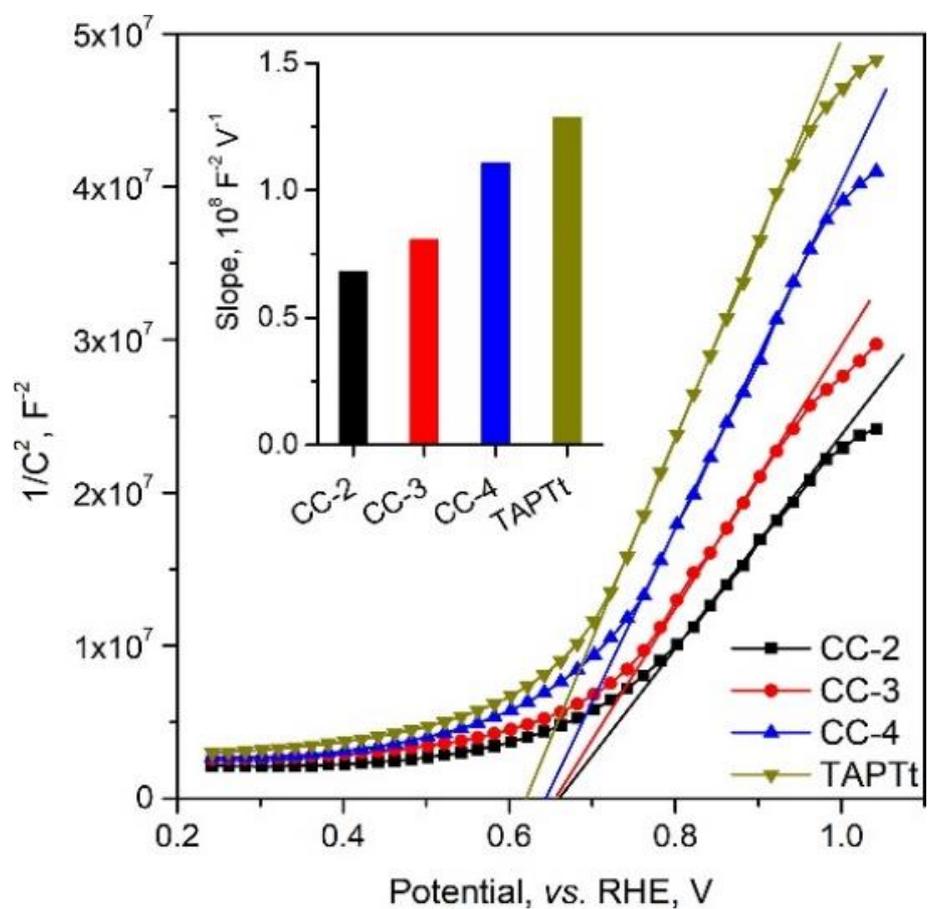

**Figure S25.** Mott-Schottky plots of TAPTt-COF and CC-x VdWHs. The fitted slopes are compared in the Inset.

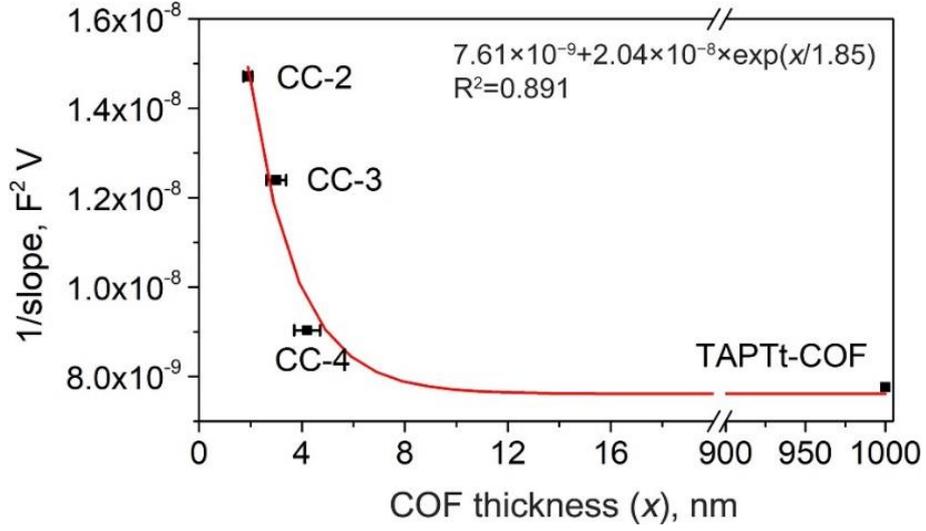

**Figure S26**. Relationship between the reciprocal of the slope in the Mott-Schottky plot as a function of the COF shell thickness. The thickness of pristine TAPTt-COF was set to 1000 nm, representing the radius of the COF spheres determined from SEM observation (**Figure S5**).

The data were fitted by solving the one-dimensional continuity equation of an unbiased n-type sample under steady-state by setting the boundary conditions as $p_n(0)$ = constant and $p_n(\infty) = p_{n0}$[21]:

$$\frac{\partial p_n}{\partial t} = 0 = -\frac{p_n - p_{n0}}{\tau_p} + D_p \frac{\partial^2 p_n}{\partial^2 x} \tag{S1}$$

The solution of Eq1 can describe the excess carrier decay in a semi-infinite semiconductor under steady-state (as exemplified in **Figure S24**), can be written as:

$$p_n(x) = p_{n0} + [p_n(0) - p_{n0}]exp\left(-\frac{x}{L_p}\right) \tag{S2}$$

where $p_n(x)$ is the carrier density function and $L_p$ is the diffusion length.[21]

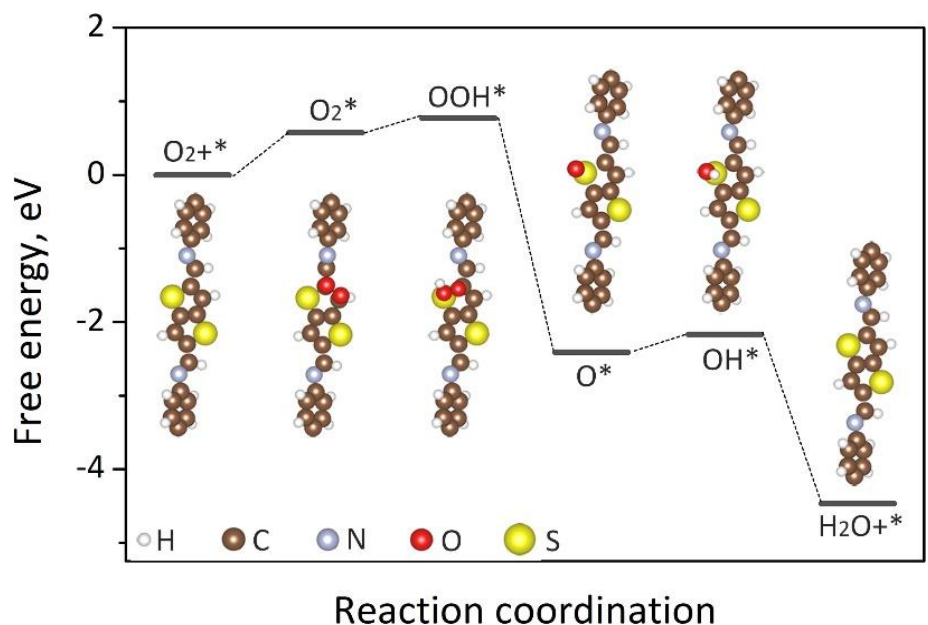

**Figure S27.** The free energy diagram of oxygen reaction intermediates on the most feasible S-C active site. Images show the corresponding optimized geometries of reaction intermediates on the active site. It should be noted that due to the inert nature of *p*-block elements and saturated COF covalent chains, the formation of OOH* and OH* is thermodynamically less favorable to that on Pt-based catalysts. The free energies of oxygen reaction intermediates are usually high. However, a previous study pointed out that the significant hydrophobicity of *p*-block elements likely leads to larger reaction prefactors[22], that make the reaction facile under experimental conditions. Meanwhile, as illustrated in **Figure 2a** in the main-text, the presence of the graphene substrate leads to a significant electron gain in COF, which may further improve its reaction activity.

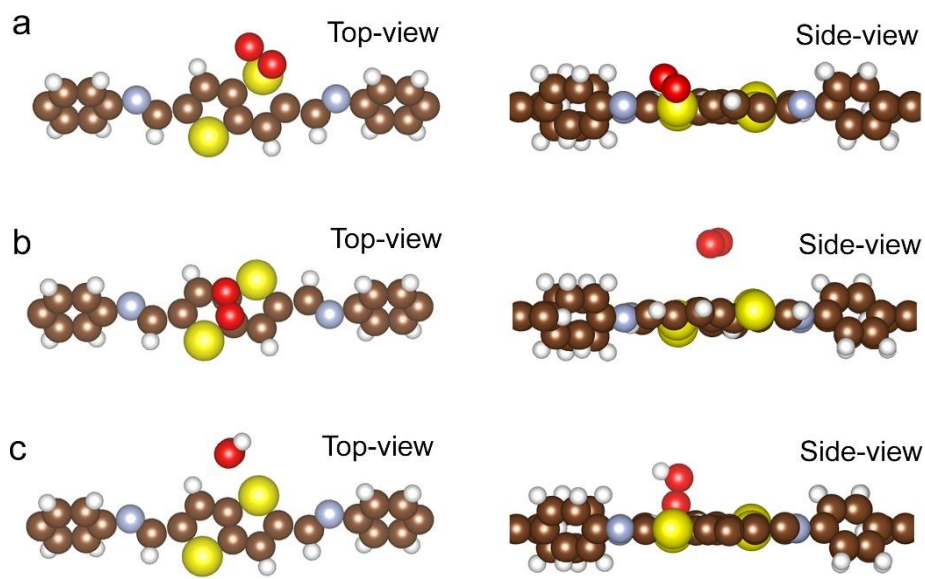

**Figure S28.** DFT models of ORR intermediates adsorbed on other possible sites on TAPTt COF. (a) The optimized geometry of the initial $O_2$ adsorption on the S site. The adsorption free energy is 2.05 eV, which forms a high energy barrier for the reaction to proceed. The adsorption of (b) $O_2$ and (c) OOH intermediates on other sites in the thienothiophene moiety results in the non-adsorption configuration, indicating that these sites are not active for ORR. The atom color scheme is the same as that in Figure S27.

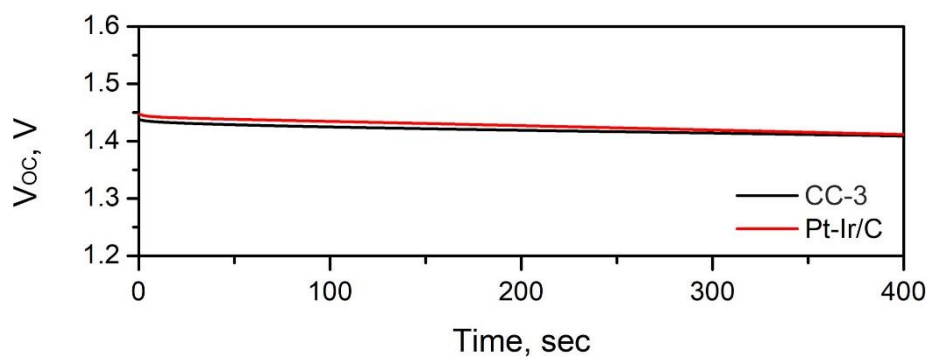

**Figure S29.** Open circuit potential vs. time of ZABs assembled using CC-3 and Pt-Ir/C catalysts.

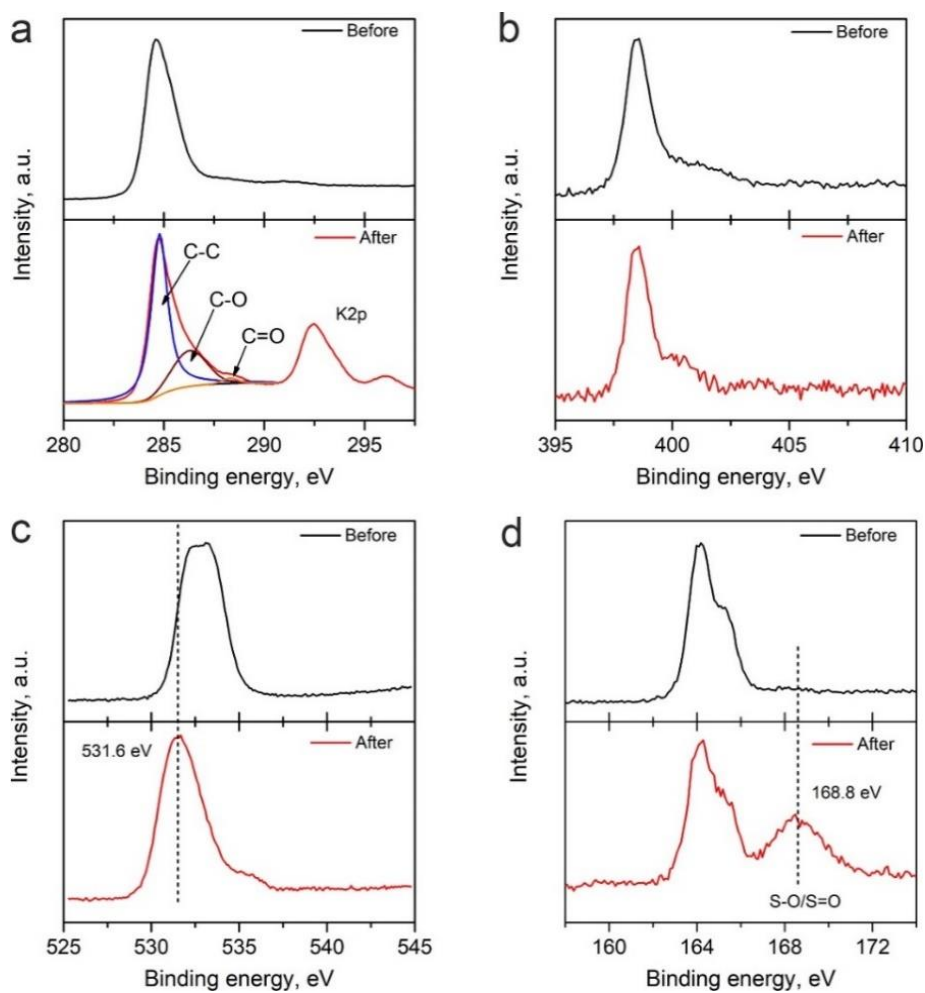

**Figure S30.** XPS spectra of (a) C, (b) N, (c) S, and (d) O in CC-3 after the ZAB cycling test over 120 discharging/charging cycles.

**References cited in the Supporting Information**

1. Guo, Y.; Yao, S.; Gao, L.; Chen, A.; Jiao, M.; Cui, H.; Zhou, Z., Boosting bifunctional electrocatalytic activity in S and N co-doped carbon nanosheets for high-efficiency Zn–air batteries. *Journal of Materials Chemistry A* **2020,** *8* (8), 4386-4395.
2. Zong, L.; Wu, W.; Liu, S.; Yin, H.; Chen, Y.; Liu, C.; Fan, K.; Zhao, X.; Chen, X.; Wang, F.; Yang, Y.; Wang, L.; Feng, S., Metal-free, active nitrogen-enriched, efficient bifunctional oxygen electrocatalyst for ultrastable zinc-air batteries. *Energy Storage Materials* **2020,** *27*, 514-521.
3. Xiao, X.; Li, X.; Wang, Z.; Yan, G.; Guo, H.; Hu, Q.; Li, L.; Liu, Y.; Wang, J., Robust template-activator cooperated pyrolysis enabling hierarchically porous honeycombed defective carbon as highly-efficient metal-free bifunctional electrocatalyst for Zn-air batteries. *Applied Catalysis B: Environmental* **2020,** *265*, 118603.
4. Li, Q.; He, T.; Zhang, Y.-Q.; Wu, H.; Liu, J.; Qi, Y.; Lei, Y.; Chen, H.; Sun, Z.; Peng, C., Biomass Waste-Derived 3D Metal-Free Porous Carbon as a Bifunctional Electrocatalyst for Rechargeable Zinc–Air Batteries. *ACS Sustainable Chemistry & Engineering* **2019,** *7* (20), 17039-17046.

5. Li, P.; Jang, H.; Zhang, J.; Tian, M.; Chen, S.; Yuan, B.; Wu, Z.; Liu, X.; Cho, J., A Metal-Free N and P-Codoped Carbon Nanosphere as Bifunctional Electrocatalyst for Rechargeable Zinc-Air Batteries. *Chemelectrochem* **2019,** *6* (2), 393-397.
6. Lin, X.; Peng, P.; Guo, J.; Xiang, Z., Reaction milling for scalable synthesis of N, P-codoped covalent organic polymers for metal-free bifunctional electrocatalysts. *Chemical Engineering Journal* **2019,** *358*, 427-434.
7. Wang, H.-F.; Tang, C.; Zhang, Q., Template growth of nitrogen-doped mesoporous graphene on metal oxides and its use as a metal-free bifunctional electrocatalyst for oxygen reduction and evolution reactions. *Catalysis Today* **2018,** *301*, 25-31.
8. Sun, T.; Wang, J.; Qiu, C.; Ling, X.; Tian, B.; Chen, W.; Su, C., B, N Codoped and Defect-Rich Nanocarbon Material as a Metal-Free Bifunctional Electrocatalyst for Oxygen Reduction and Evolution Reactions. *Advanced Science* **2018,** *5* (7), 1800036.
9. Jia, N.; Weng, Q.; Shi, Y.; Shi, X.; Chen, X.; Chen, P.; An, Z.; Chen, Y., N-doped carbon nanocages: Bifunctional electrocatalysts for the oxygen reduction and evolution reactions. *Nano Research* **2018,** *11* (4), 1905-1916.
10. Qian, Y.; Hu, Z.; Ge, X.; Yang, S.; Peng, Y.; Kang, Z.; Liu, Z.; Lee, J. Y.; Zhao, D., A metal-free ORR/OER bifunctional electrocatalyst derived from metal-organic frameworks for rechargeable Zn-Air batteries. *Carbon* **2017,** *111*, 641-650.
11. Liu, S.; Zhang, H.; Zhao, Q.; Zhang, X.; Liu, R.; Ge, X.; Wang, G.; Zhao, H.; Cai, W., Metal-organic framework derived nitrogen-doped porous carbon@ graphene sandwich-like structured composites as bifunctional electrocatalysts for oxygen reduction and evolution reactions. *Carbon* **2016,** *106*, 74-83.
12. Lu, Z.; Wang, J.; Huang, S.; Hou, Y.; Li, Y.; Zhao, Y.; Mu, S.; Zhang, J.; Zhao, Y., N, B-codoped defect-rich graphitic carbon nanocages as high performance multifunctional electrocatalysts. *Nano energy* **2017,** *42*, 334-340.
13. Gao, R.; Dai, Q.; Du, F.; Yan, D.; Dai, L., C60-adsorbed single-walled carbon nanotubes as metal-free, pH-universal, and multifunctional catalysts for oxygen reduction, oxygen evolution, and hydrogen evolution. *Journal of the American Chemical Society* **2019,** *141* (29), 11658-11666.
14. Yang, H. B.; Miao, J.; Hung, S.-F.; Chen, J.; Tao, H. B.; Wang, X.; Zhang, L.; Chen, R.; Gao, J.; Chen, H. M., Identification of catalytic sites for oxygen reduction and oxygen evolution in N-doped graphene materials: Development of highly efficient metal-free bifunctional electrocatalyst. *Sci. Adv.* **2016,** *2* (4), e1501122.
15. Liu, Q.; Wang, Y.; Dai, L.; Yao, J., Scalable fabrication of nanoporous carbon fiber films as bifunctional catalytic electrodes for flexible Zn-air batteries. *Advanced Materials* **2016,** *28* (15), 3000-3006.
16. El-Sawy, A. M.; Mosa, I. M.; Su, D.; Guild, C. J.; Khalid, S.; Joesten, R.; Rusling, J. F.; Suib, S. L., Oxygen Reactions: Controlling the Active Sites of Sulfur-Doped Carbon Nanotube–Graphene Nanolobes for Highly Efficient Oxygen Evolution and Reduction Catalysis (Adv. Energy Mater. 5/2016). *Advanced Energy Materials* **2016,** *6* (5).
17. Qu, K.; Zheng, Y.; Dai, S.; Qiao, S. Z., Graphene oxide-polydopamine derived N, S-codoped carbon nanosheets as superior bifunctional electrocatalysts for oxygen reduction and evolution. *Nano Energy* **2016,** *19*, 373-381.
18. Zhang, J.; Zhao, Z.; Xia, Z.; Dai, L., A metal-free bifunctional electrocatalyst for oxygen reduction and oxygen evolution reactions. *Nature nanotechnology* **2015,** *10* (5), 444-452.
19. Tian, G. L.; Zhao, M. Q.; Yu, D.; Kong, X. Y.; Huang, J. Q.; Zhang, Q.; Wei, F., Nitrogen-doped graphene/carbon nanotube hybrids: in situ formation on bifunctional catalysts and their superior electrocatalytic activity for oxygen evolution/reduction reaction. *Small* **2014,** *10* (11), 2251-2259.
20. Ren, F.; Kanaan, S. A.; Majewska, M. M.; Keskar, G. D.; Azoz, S.; Wang, H.; Wang, X.; Haller, G. L.; Chen, Y.; Pfefferle, L. D., Increase in the yield of (and selective synthesis of large-diameter) single-walled carbon nanotubes through water-assisted ethanol pyrolysis. *Journal of Catalysis* **2014,** *309*, 419-427.
21. Sze, S. M.; Ng, K. K., Physics and Properties of Semiconductors—A Review. In *Physics of Semiconductor Devices*, John Wiley & Sons: 2006; pp 5-75.
22. Tripkovic, V.; Vegge, T., Potential- and Rate-Determining Step for Oxygen Reduction on Pt(111). *The Journal of Physical Chemistry C* **2017,** *121* (48), 26785-26793.